# Optimizing image capture for low-light widefield quantitative fluorescence microscopy

Zane Peterkovic ✉ ; Avinash Upadhya ; Christopher Perrella ; Admir Bajraktarevic ;
Ramses E. Bautista Gonzalez ; Megan Lim ; Kylie R. Dunning ; Kishan Dholakia ✉

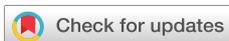

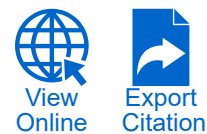

## Articles You May Be Interested In

Aberration control in quantitative widefield quantum microscopy

*AVS Quantum Sci.* (September 2022)

High contrast, depth-resolved thermoreflectance imaging using a Nipkow disk confocal microscope

*Rev. Sci. Instrum.* (January 2010)

Two-dimensional spatiotemporal focusing of femtosecond pulses and its applications in microscopy

*Rev. Sci. Instrum.* (August 2015)

# Optimizing image capture for low-light widefield quantitative fluorescence microscopy

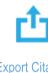

View Online   Export Citation   CrossMark

Zane Peterkovic,[1,a)] Avinash Upadhya,[1] Christopher Perrella,[1] Admir Bajraktarevic,[1] Ramses E. Bautista Gonzalez,[1] Megan Lim,[2] Kylie R. Dunning,[2] and Kishan Dholakia[1,3,a)]

AFFILIATIONS

[1] Centre of Light for Life and School of Biological Sciences, University of Adelaide, Adelaide, South Australia 5005, Australia
[2] Centre of Light for Life and The Robinson Research Institute, School of Biomedicine, University of Adelaide, Adelaide, South Australia 5005, Australia
[3] SUPA, School of Physics and Astronomy, University of St Andrews, North Haugh, Fife KY16 9SS, United Kingdom

[a)] Authors to whom correspondence should be addressed: kishan.dholakia@adelaide.edu.au
and zane.peterkovic@adelaide.edu.au

## ABSTRACT

Low-light optical imaging refers to the use of cameras to capture images with minimal photon flux. This area has broad application to diverse fields, including optical microscopy for biological studies. In such studies, it is important to reduce the intensity of illumination to reduce adverse effects such as photobleaching and phototoxicity that may perturb the biological system under study. The challenge when minimizing illumination is to maintain image quality that reflects the underlying biology and can be used for quantitative measurements. An example is the optical redox ratio, which is computed from autofluorescence intensity to measure metabolism. In all such cases, it is critical for researchers to optimize the selection and application of scientific cameras to their microscopes, but few resources discuss performance in the low-light regime. In this tutorial, we address the challenges in optical fluorescence imaging at low-light levels for quantitative microscopy, with an emphasis on live biological samples. We analyze the performance of low-light scientific cameras including electron-multiplying charge-coupled device, scientific complementary metal oxide semiconductor (sCMOS), and the photon-counting sCMOS architecture, termed quantitative CMOS, while considering the differences in platform architecture and the contribution of various sources of noise. The tutorial covers a detailed discussion of user-controllable parameters, as well as the application of post-processing algorithms for denoising. We illustrate these concepts using autofluorescence images of live mammalian embryos captured with a two-photon light sheet fluorescence microscope.

## I. INTRODUCTION

Over the course of the past century, fluorescence microscopy has emerged as a powerful tool across various scientific disciplines. In biology, it has played an important role in studying subcellular structures and processes. The first studies focused on autofluorescent specimens; however, the advent of fluorescent stains revolutionized the field, enabling the visualization of specimens that were hitherto unobservable. Subsequently, the development of fluorescently labeled antibodies opened new avenues for visualizing specific targets within cells. A major breakthrough occurred in the 1960s with the discovery of green fluorescent protein (GFP) in jellyfish, followed by the successful cloning of the gene encoding GFP in 1992. These advancements transformed fluorescence microscopy and scientific research by enabling the use of GFP as a targeted genetic tag, allowing real-time tracking of proteins within live cells. This significant advancement earned its pioneers a Nobel Prize in 2008.[1]

As fluorescence microscopy has evolved, innovative markers have spurred advancements in the field, allowing for high-resolution studies of subcellular dynamics across both spatial and temporal domains. This progress has been fueled by interdisciplinary collaboration enabling biologists to leverage this powerful technique. Chemists have contributed by developing novel fluorescent probes, while physical scientists have advanced the hardware and software of the microscopy platforms.

In recent years, biologists have been increasingly drawn to label-free optical imaging due to its ability to study live cells and tissues in their natural state without the potential interference of fluorescent dyes or tags. This approach also enables the study of downstream biological processes with minimal interference, offering insights into how initial cellular events impact subsequent development. Label-free optical imaging relies on endogenous cellular fluorescence, commonly termed autofluorescence, which often leads to low-light levels from the sample. This presents challenges for fluorescence imaging systems to enhance their detection sensitivity. Recent studies utilizing autofluorescence have demonstrated its efficacy in identifying cancerous cells,[2,3] assessing metabolic health of pre-implantation embryos[4–7] and monitoring the health of organoids.[8]

Fluorescence imaging relies on illumination provided by laser beams or other high-intensity sources, which may cause photobleaching and photodamage.[9] To preserve cell viability, there is a major motivation to reduce the incident radiation, while sustaining sufficient emission to record an appropriate signal. With these points in mind, the light sheet fluorescence microscope (LSFM) has emerged to mitigate these concerns.[9–12] LSFM is a powerful wide-field modality, able to reveal three-dimensional information from a sample via plane-by-plane imaging. In this tutorial, we use this modality together with the pre-implantation mouse embryo to demonstrate key aspects to consider for low-light level wide-field imaging for the biosciences.

Wide-field fluorescence microscopy involves the excitation of fluorescence across a sample volume, which is then collected by an imaging array, avoiding the lengthy scan times of point-scanning microscopy methods (e.g., confocal microscopy).[13,14] Scientific cameras serve as major components in all wide-field imaging systems and often place limits on what one may faithfully capture with a given microscope. Low-light imaging is one such major limitation: as the intensity incident on the camera array decreases, fluctuations in pixel value ("camera noise") induced by internal electronics grow to predominate. This leads not only to degradation in the qualitative appearance of the image but also possible misinterpretation of data when used in quantitative analysis.[15–17] This is especially a concern for autofluorescence imaging, where vital molecular information is extracted through the quantitative imaging of weak fluorescence.[18–21]

Despite its major importance, the details and optimal use of scientific cameras remain opaque for many researchers: it is often unclear which imaging platform is best suited for a given application or how to best optimize the imaging platforms currently in use. Furthermore, recent developments in high-end camera architectures for low-light imaging have not been fully explored for fluorescence microscopy.[16,22,23] These point toward a need for an assessment of modern imaging technology and how to optimize it for biological imaging.

In this tutorial, we perform a comparison of different specialized camera architectures for low-light imaging, which will be discussed in Sec. I. Previous analyses of scientific cameras for biophotonics have typically focused on how to assess these cameras for noise statistics, without direct connection to the practicalities of biological imaging.[15,17,24,25] In addition, consideration has not been given to specialized camera modes such as "confocal line scan," also known as light sheet mode, which can be used in particular circumstances to reduce background noise.[26] This tutorial provides a resource for researchers interested in optimizing the performance of their wide-field imaging systems for low-light level imaging by providing information relevant to scientific camera operation for biological imaging. This includes the origin and nature of image noise, the effect of controllable parameters such as exposure time and pixel size on image capture, and the use of machine learning algorithms for post-processing denoising, including their pitfalls. Through this discussion, we also provide methods for the assessment of camera noise statistics and the empirical determination of fluorescence image quality—these approaches are of value for those researchers seeking to assess the performance of their own imaging platforms.

This tutorial is structured as follows: Sec. II is divided in two, with Sec. II A covering fundamental aspects of camera operation such as digitization and quantum efficiency, while Sec. II B explains the nature and origin of noise present during the imaging process, including a model of camera noise as a function of incident intensity. This is followed by Sec. III, where the insights of Sec. II are applied to real-world data captured using a wide-field fluorescence microscope. Section III A discusses the difficulties in quantifying image quality, while introducing the empirical approach used in this tutorial, while Secs. III B–III F analyze the effects of camera architecture, pixel size, exposure time, and operating modes for the purpose of optimizing image capture. Section III G introduces the use of emerging deep learning algorithms for the purpose of denoising fluorescence images. We conclude with a discussion on the numerous difficulties and trade-offs attendant to fluorescence imaging with scientific cameras.

In order to be representative of the current state of the art in imaging technology, three cameras have been selected for comparison: the Hamamatsu Photonics ORCA-Flash V3 (C13440-20CU), the more advanced Hamamatsu Photonics ORCA-Quest (C1550-20UP), both based on scientific complimentary metal–oxide–semiconductor (sCMOS) platform, and the Andor Technology iXON Life 888, based on an electron-multiplying charge-coupled device (EMCCD) platform. Of these, the ORCA-Quest and iXON Life represent high-end imaging platforms with specialized camera architectures for low-light image capture (see Sec. I for a discussion on camera architecture), while the ORCA-Flash is less specialized, being an effective mid-range scientific camera represented among groups seeking to capture autofluorescent emission.[20,27–29] It should be noted that there exist a number of other wide-field, low-light imaging technologies such as the intensified CCD (ICCD), wide-field microchannel plate detector, and single photon avalanche diode (SPAD) array. These technologies are primarily used for extremely high-speed, high time resolution imaging; while there are biological applications of these technologies,[30–32] this tutorial focuses on the more ubiquitous forms of scientific camera available. Of these technologies, the ICCD has similarities to the EMCCD, possessing similar architecture; certain results of this tutorial will, therefore, apply to ICCDs.

The *in situ* measurements for this tutorial were captured with a purpose-built two-photon LSFM, details for which are found in Appendix B 1. LSFM is an increasingly popular wide-field imaging geometry due to its minimal photodose,[9,10,12] especially for the imaging of autofluorescent samples.[4,6,20,27–29,33] Moreover, the flexibility of LSFM has also enabled quantitative and computationally intensive imaging modalities, such as super-resolution techniques.[34,35] The system used in this tutorial represents a case where the camera itself forms a good platform for testing capabilities for quantitative fluorescence microscopy.

While our analysis has been based on the cameras and the chosen optical system, namely, LSFM, the insights are applicable to wide-field fluorescence microscopy generally. Indeed, the analysis of camera noise statistics is broadly consistent across disciplines for applications outside biological imaging.[36]

## II. SCIENTIFIC CAMERA FUNDAMENTALS

### A. Camera operation fundamentals

Optical microscopy is a powerful technique that allows for the observation of microscopic to mesoscopic fields of view using light, finding notable application in fields such as materials science, chemistry, and biology. Although microscope designs are diverse, the scientific cameras they employ are based on similar principles: digital measurement of light intensity across an array of photodetectors. In the following sections, we discuss fundamental aspects of scientific cameras such as different sensor architectures in Sec. II A 1. The process of how an image is captured is described in Secs. II A 1–II A 3. Following this in Sec. II B, we discuss how the presence of noise in the detected image can deteriorate image quality.

#### 1. Camera architecture

Before describing the imaging process, it is generally necessary to briefly introduce digital camera architecture, as it is especially relevant to low-light imaging. Digital camera sensors come in two types: complementary metal–oxide–semiconductor (CMOS) pixel arrays, which are the ones used in the scientific CMOS (sCMOS) ORCA-Flash and ORCA-Quest, and second, the charge-coupled device (CCD) such as the iXON Life, this being a specialized electron-multiplying CCD (EMCCD). The major distinction between these architectures is the location of the read-out electronics, where the charge of each pixel is converted to a voltage: from Fig. 1(a), we can see that for a CMOS camera, the read-out occurs at the site of the pixel; meanwhile, Fig. 1(b) shows that CCD sensors first transport the charges to the read register, where they are then read out at the end of the frame.

This difference is significant as the degradation of the image associated with this read-out, the "read noise" is one of the most significant noise contributions in low-light imaging, discussed in detail in Sec. II B. The design of CCD architectures allows for an electron-multiplication process to occur prior to read-out by the addition of an extended read register, as can be seen in Fig. 1(c). This electron-multiplication process decreases the effect of the read noise upon the pixel measurement, but this is at the expense of increased "shot noise"[37] (this "trade-off" is discussed in Sec. II B). These EMCCD cameras have been used in low-light applications, such as photon counting.[38]

CMOS sensor technology is a more recent development when compared to CCD sensor technology.[39] There was a period of maturation for several years before this technology found scientific use.[40] CMOS cameras used for scientific applications are dubbed scientific CMOS or sCMOS cameras. They have found widespread application, including in demanding fluorescence microscopy applications such as single-molecule localization microscopy.[41] With the introduction of the specialized low-noise sCMOS architecture, termed quantitative CMOS (qCMOS), comparable levels of sensitivity and photon-counting capabilities to EMCCD cameras have been achieved with the CMOS platform.

The relative performance of these different camera types is an area of great interest and helps inform the user as to which camera to choose for a given application. Comparisons between qCMOS and EMCCD cameras have demonstrated the potential for the former in quantum experiments.[42] Previous comparisons between sCMOS and EMCCD cameras for the purpose of bioimaging neither considered these recent developments in camera technology nor focused on specific cases in biophotonics, especially low-light imaging.[16,22,23] Therefore, to date, there is no suitable comparison between sCMOS, qCMOS, and EMCCDs for low-light fluorescence microscopy in the literature, which is a gap that this tutorial bridges.

#### 2. Detection of light

Regardless of digital camera architecture, the process of detecting photons and converting them into a digital image is common across the platforms. This is depicted in Fig. 2. Each pixel of a camera sensor consists of a photodiode which, during imaging, is "active" for a given "exposure time." During the exposure time, incident photons can generate photo-electrons (e−), which accumulate as a charge within the photodiode. The fraction of photons that are converted into photo-electrons, expressed as a percentage, is referred to as the quantum efficiency (QE)[36] of the pixel or sensor. QE arises from quantum mechanical effects and is dependent on the wavelength of incident light.

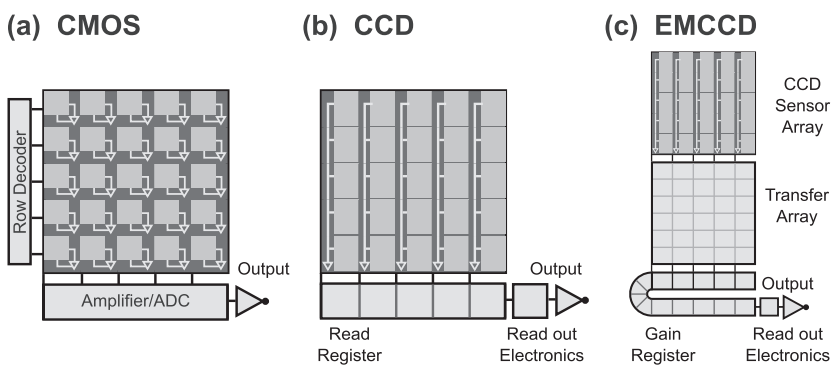

**FIG. 1.** Diagram of different scientific camera architectures. (a) CMOS sensor array, depicting on-pixel read-out electronics and separate row/column electronics. (b) CCD array, depicting how charges are transported to the read register prior to read-out. (c) EMCCD architecture, including transfer array and gain register.

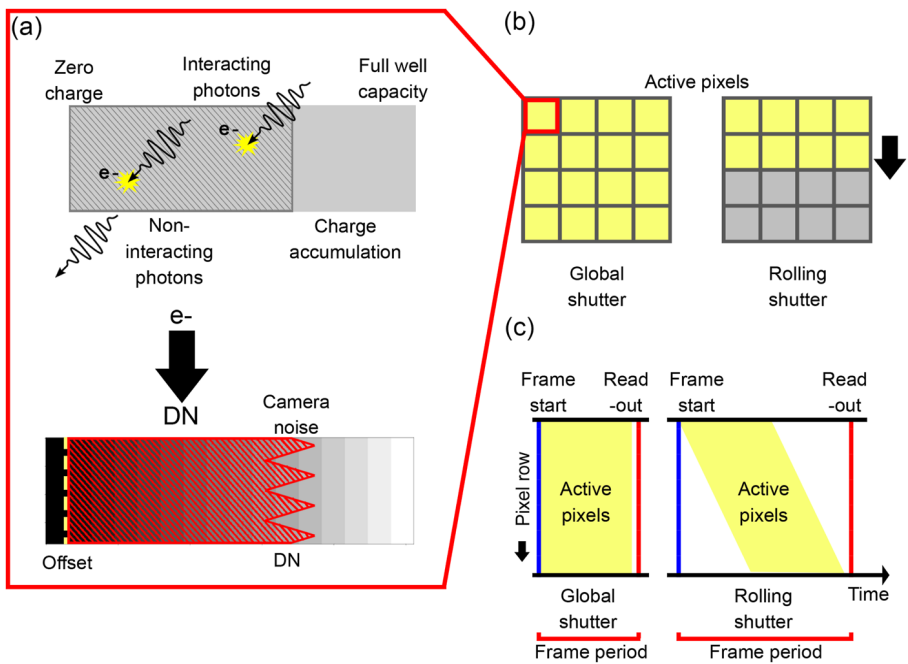

**FIG. 2.** Conversion process of a digital camera, from detection to frame read-out. (a) Diagram of charge accumulating on a pixel: interacting photons generate a build-up of photo-electrons "filling up" a pixel. The accumulated charge is then converted to a digital number (DN, see later) based on the set bit-depth; the offset is highlighted, which serves as the "zero" read out for a pixel. (b) Global shutter compared to a rolling shutter. With the global shutter, all pixel rows are active at the same time, whereas for a rolling shutter, the pixels "cascade" from the top; all pixels are active for the same exposure time, but not synchronously. It should be noted that the frame period is slightly longer than the active pixel time. (c) "Rolling shutter" read out as opposed to "global shutter" read out when considering the time trace of the active pixels.

This conversion of photons into photo-electrons only happens within the active area of the pixel, which is not typically the full area of the pixel. The "fill-factor" quantifies the ratio between the active to non-active surface area on the pixel and is a fixed quantity for a particular sensor design. A higher QE and "fill-factor" leads to efficient capture of light, which is highly desirable in low-light imaging. Distinguishing between the effects of QE and fill-factor is difficult, and manufacturers do not always clearly delineate between them. As it is challenging to infer the presence of non-interacting photons, we will be considering the fill-factor as part of the QE, and the intensities in this tutorial will be reported as "detected photons" or "photo-electrons."

Different cameras exhibit different QEs across the optical spectrum. For example, the iXON Life 888 has a QE of <90% in the spectral range 500–700 nm, but <60% at 400 nm.[43] Typically, modern silicon-based scientific cameras have high QEs in the center of the visible range, with diminishing responsivity into the ultraviolet and infrared, with CCD cameras having a slight edge over their counterparts. Figure 3(a) depicts a comparison between the QE for different cameras across the visible spectral range. While the ORCA-Flash and iXON Life have similar profiles, the ORCA-Quest is distinctively skewed toward attaining a higher QE at shorter wavelengths.

The importance of varying QEs across the optical spectrum to quantitative microscopy can be illustrated by considering nicotinamide adenine dinucleotide (phosphate) [NAD(P)H] and flavin adenine dinucleotide (FAD). These are two prominent metabolic co-enzymes and endogenous fluorophores.[44,45] Figure 3(b) depicts the effect of camera QE on the normalized intensity spectra of these fluorophores.[44] The measurements of NAD(P)H and FAD autofluorescence are used to calculate the optical redox ratio, a ratiometric calculation of autofluorescence, which can act as a marker of metabolic state.[21] The optical redox ratio shows relative changes in the oxidation–reduction state in the cell and has been used to assess embryo viability.[4,7,46] Figure 3(b) depicts the effect that different camera QE curves can have on the detected spectrum; the ORCA-Quest and ORCA-Flash are approximately equivalent for the FAD signal, whereas the ORCA-Flash reports ∼20% lower intensity for NAD(P)H. As such, the effect of camera QE may be overlooked in the measurement of optical redox ratio through fluorescence microscopy.

During the exposure time of a pixel, the number of photo-electrons produced, $S^{e-}$, for a pixel of area, $A_{px}$, and exposure time, $t_{exp}$, can be expressed as

$$S^{e-} = QE \cdot A_{px} \cdot t_{exp} \cdot I, \quad (1)$$

where $I$ is the intensity of the light falling on the pixel, in units of photons/$\mu m^2$/s. Ultimately, it is $I$ that is the quantity of physical interest, so all factors in Eq. (1) must be considered when performing quantitative microscopy.

Over the exposure time of a pixel, the number of photo-electrons able to be accumulated is finite and is limited by the "full well depth" of the pixel. Once the full well depth is reached, any further incident photons will fail to register, and the pixel is deemed to be "saturated" or "overexposed." When imaging heterogeneous samples, it is likely that only certain pixels will be saturated, meaning that the relative intensities reported across the field of view do not reflect the underlying ground truth, which can be detrimental for quantitative analysis. Although overexposure is not typically a concern for low-light imaging, long exposure times can lead to pixel saturation even with low-intensity light and thus should be taken into consideration.

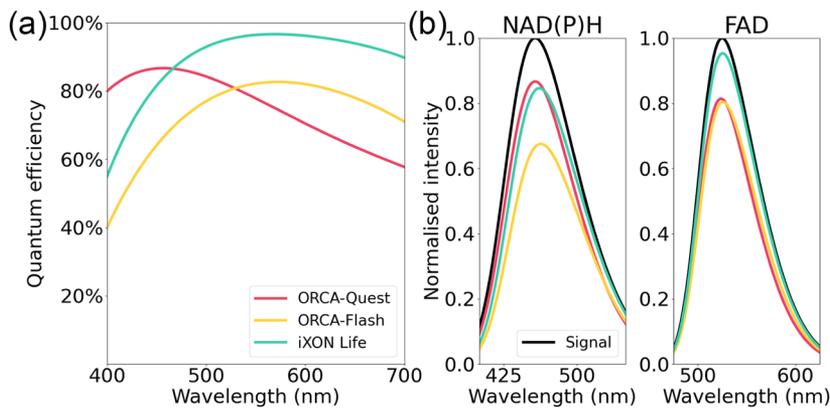

**FIG. 3.** Quantum efficiency curves and their effect on the observed spectra. (a) Quantum efficiency curves for the cameras considered in this tutorial, derived from the data provided via camera technical specifications. (b) Effect of quantum efficiency on the normalized intensity spectra for NAD(P)H and FAD when excited by 730 nm two-photon emission. These spectra are adapted from Huang *et al.*[44]

### 3. Pixel and frame read-out

At the end of the set exposure time, the number of photo-electrons per pixel must be converted to a value that can be stored digitally. A full description of this process is outside the scope of this tutorial [see Janesick for further details[36]]. Here, it is sufficient to highlight two major steps: read-out and analog-to-digital conversion, common to both CMOS and CCD cameras. Read-out refers to the conversion of accumulated charge in photo-electrons to a voltage, the magnitude of which is converted to a digital number (DN), also called a "count."

This process is governed by the conversion gain $g$ (with units DN/e−), or equivalently its inverse, the "analog-to-digital conversion factor" $\frac{1}{g} = K_{ADC}$(e−/DN). This conversion gain is the product of numerous sources of gain internal to the camera electronics,[36] some of which are fixed and some of which can be set by the user. Typically the conversion gain $g$ is reported for a given bit-depth $g_n$, and varying the bit-depth changes it according to $g_m = g_n\left(\frac{2^n}{2^m}\right)$, where $n$ and $m$ are the bit-depths of the ADC.

One notable form of user-controllable gain is the electron-multiplication gain exclusive to EMCCD cameras. As mentioned in Sec. I, the architecture of CCD cameras allows for this unique gain process, and with this gain enabled $g$ is dramatically increased. While this electron-multiplication is unique to EMCCD cameras, cameras of any architecture may offer a controllable gain, and any change will have consequences for the conversion of photo-electrons into DNs. It is worth mentioning that there exist methods of empirically determining $g$ based on uniform illumination, a description of which is provided in Appendix A 3.

Finally, one aspect of the ADC that should not be overlooked is the addition of an "offset." During the digitization procedure, the offset serves as the "zero value" for the digital numbers, which prevents stack-overflow. As such, the offset contributes a non-physical addition to the DN for a given pixel; while this contribution is usually negligible, it must be considered when performing low-light quantitative microscopy.

The discussion so far has applied to the conversion process for a single pixel. In order for an image to be captured, the entire array of pixels must be read out, which we refer to as a frame. There are two common approaches to read-out the full frame in a modern digital camera: "rolling shutter" mode and "global shutter" mode, which are depicted in Figs. 2(b) and 2(c). For the global shutter method, every pixel in the array begins and ends its exposure time at the same time; with the rolling shutter method, the active pixels are "staggered" by row or column, meaning that the same moment in time is not necessarily being imaged within the same frame. Rolling shutter is found primarily among sCMOS cameras, with some select few offering the ability to select the shutter method. While a rolling shutter is perfectly suited to many applications, moving samples can lead to artifacts and distortions in the captured frame.

After digitization of the full camera frame, the process can start again to record the next image. The rate at which the camera captures frames in time is referred to as the frame rate, and is reported in Hertz (Hz) or, equivalently, frames per second (fps). The reciprocal of this is the "frame period." For certain applications, such as imaging dynamic processes, these are parameters of major interest. The (user-controllable) exposure time, along with the pixel read-out time, sets a lower limit on the frame period, and, therefore, limits the frame rate.

This "read-out time" is highly dependent on a number of factors, including camera architecture, data transfer method, array size, and bit-depth. Generally, sCMOS cameras have higher frame-rates due to the on-pixel read-out electronics, but under most circumstances, the data transfer or exposure time will be a larger bottleneck.

Even under ideal conditions, achieving frame-rates of >250 Hz is challenging, effectively limiting the temporal resolution to <5 ms. Applications that require finer temporal resolution may require time-gating electronics or high-speed imaging technology, such as ICCD or SPAD cameras.[31]

In this section, we have discussed how a camera is able to convert light incident on its sensor into arrays or "frames" of DNs. The output digital number for the pixel $i$ for the frame $F$ can be expressed mathematically as $DN_{i,F}(S^{e-})$.

In the ideal case, free of camera noise, the relationship is as follows:

$$\text{DN}_{i,F} = S^{e-}_{i,F} g + \text{Off}_i, \quad (2)$$

where $\text{Off}_i$ refers to the offset for that pixel.

Due to noise, however, there are additional factors that disrupt the relation in Eq. (2). This camera noise will be the subject of Sec. II B.

### B. Camera noise in the imaging process

The process of converting a light field into an array of digital numbers is described in Sec. II A. This, however, is not a perfect process, and there are numerous ways in which noise sources can distort the digital image produced; Fig. 4 contains a schematic of the noise sources in the photoelectron conversion procedure. It is useful to separate this noise into two types: temporal and spatial. Temporal noise refers to the discrepancy in values reported by the same pixel when exposed to identical conditions and is discussed in Sec. II B 1. The literature on camera temporal noise is more developed than that of spatial noise, and the contribution of the independent sources of temporal noise is often generically called "total noise,"[15,36] a convention we continue in this tutorial. Spatial noise refers to discrepancies of values between different pixels under identical conditions and is discussed in Sec. II B 2.

#### 1. Temporal noise

Temporal noise is of major importance as it determines the level of confidence that a reported DN reflects the actual incident intensity. Temporal noise is the result of random fluctuations from predominantly three independent noise sources: noise associated with the read-out electronics (read noise), noise associated with the detected photons (shot noise), and noise associated with the sensor array (fixed-pattern noise, FP noise). How these noise sources feed into the signal processing chain is depicted in Fig. 4.

Temporal noise has a mean contribution of 0. This means that over successive measurements, the mean of the result should correspond to the ground truth. If we consider a series of frames captured of an invariant scene, then the mean digital number for a pixel $i$ is given by

$$\overline{\text{DN}}_i = \text{Avg}_F[\text{DN}_{i,F}] = S^{e-}_i g + \text{Off}_i, \quad (3)$$

where $S^{e-}_i$ and $\text{Off}_i$ are assumed to be constant. The spread of values for this mean DN can be characterized by the standard deviation $\sigma_{\text{Total}}$, to which all sources of temporal noise contribute.

Regardless of architecture, the conversion process begins with detection, and it is this stage where shot noise enters, as depicted in Fig. 4. The excitation of photo-electrons is an inherently statistical process following a Poissonian distribution.[36] As such, there is an inherent source of temporal noise, one which cannot be eliminated save for non-classical, quantum imaging techniques, not readily applicable to fluorescence microscopy.[47,48] Furthermore, if an EMCCD camera has its electron-multiplication enabled, then photo-electrons are passed through the "gain register" [Fig. 4(b)], which has the effect of multiplying the shot noise contribution by as much as a factor of $\sqrt{2}$, known as the "excess noise factor."[37]

Putting these together, the shot noise variance in units of DN is given by

$$\sigma^2_{\text{DN,shot}} = F_n S_{e-} g^2, \quad (4)$$

where $F_n = 1$ for sCMOS cameras.

While shot noise is inherent to the detection process, read noise is an unwanted consequence of the read-out process and, therefore, is solely due to the camera. Read noise is a limiting factor on what a camera is able to capture, as it largely determines the "noise floor" beneath which excited photo-electrons are lost in the fluctuations. It is the electron-multiplication procedure of EMCCDs, which seeks to amplify the signal prior to read-out, thereby reducing the effect of read noise; sCMOS cameras read-out on the site of the pixel [Fig. 4(a)], meaning this amplification approach is not available on this platform. The development of increasingly quiet CMOS read-out electronics was an important factor in the increased use of sCMOS cameras,[40] and the modern development of qCMOS cameras with photon-counting capabilities;[49] in order for a single photon to be detected on a pixel, the read noise must be <1 e−, which has been enabled by modern advances in imaging. While EMCCD cameras still have unattainable low read noise levels compared to sCMOS cameras, they also unavoidably include the excess noise factor, which increases the effect of shot noise.

As a source of noise originating within the camera, read-noise is independent of incident signal and is, therefore, determined entirely by the camera, being a significant figure of merit,

$$\sigma_{\text{DN,read}} = g \cdot \sigma_{e-,\text{read}}. \quad (5)$$

However, there is another source of noise relevant only to sCMOS cameras: FP noise possesses both a temporal and a spatial component—fluctuations in CMOS electronics create temporal discrepancies, which are disproportionately localized to certain rows and columns, creating a patterned appearance.[15,36] In the context of this tutorial, "FP noise" will exclusively refer to the temporal contribution, while "structured noise" will refer to patterns that arise due to spatial noise. FP noise has a similar dependence on the incident light as shot noise,

$$\sigma^2_{\text{DN,FP}} = A_{\text{FP}} S_{e-}, \quad (6)$$

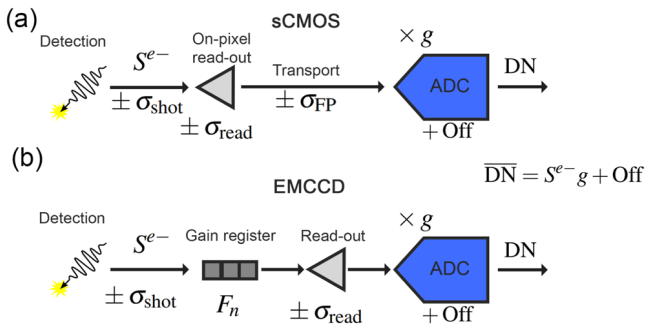

**FIG. 4.** Schematic diagrams of the conversion procedures for both sCMOS and EMCCD cameras, including sources of noise. (a) Charges on sCMOS pixels are immediately read out, leading to read-noise—the transport to the ADC introduces FP noise, while the digital conversion includes the addition of the offset. (b) If the electron-multiplying gain is enabled for the EMCCD, charges will be shifted into the grain register where they are amplified prior to read out; it is the fact that this occurs before read-out, which both introduces the excess noise factor $F_n$ while reducing the effective read noise. For both cases, the total noise is given by $\sigma^2_{\text{total}} = \sigma^2_{\text{shot}} + \sigma^2_{\text{read}} + \sigma^2_{\text{FP}}$.

where $A_{FP}$ is the "FP coefficient."

The aforementioned sources of noise are each independent, which implies that the total noise variance is equal to the sum of the variance of these different sources of noise,

$$\sigma^2_{DN, total} = \sigma^2_{DN, shot} + \sigma^2_{DN, FP} + \sigma^2_{DN, read}. \quad (7)$$

When we take this relationship and the expressions for the noise standard deviations above, the following total noise function is able to be derived:

$$\sigma_{DN, total}(S_{e-}) = \sqrt{F_n S_{e-} g^2 + A_{FP} S_{e-} + \sigma^2_{DN, read}}. \quad (8)$$

Equation (8) shows that given a select number of parameters ($F_n, g, A_{FP}, \sigma_{DN,read}$), we are able to predict the temporal noise for the average camera pixel as a function of interacting photons; all of which are able to be derived with a photon transfer curve (see Appendix A 3). Figure 5(a) contains an example photon transfer curve for the ORCA-Quest, with empirical measurements for the various contributions by the independent noise sources; Fig. 5(b) demonstrates the relative predominance of each noise source. What is clear is that shot noise makes by far the largest contribution, with read-noise only dominating in the low-light regime, and FP noise never adding a dominant contribution.

Figure 5(c) compares the total noise functions of the cameras under analysis; most apparent are the differences between the "noise floors," imposed by the read noise of the various cameras. Outside of this low-light regime, they exhibit remarkably similar behaviors.

It should be noted that this function describes only camera noise—before conversion even begins, the pixel size, exposure time, and quantum efficiency determine how many photons are detected, regardless of their origin. Fluorescent emission is incoherent, and especially when dealing with thick or highly scattering samples, a background of scattered photons will also be present in the image.

How this is accounted for in our analysis will be discussed in Sec. III A.

There is another source of noise which is frequently discussed in the context of scientific cameras: this is the dark signal, otherwise termed the dark noise. This is noise which originates in the thermal excitation of electrons within the camera, leading to distortions or other false events. Modern scientific cameras are cooled to reduce the effect of dark noise on the imaging process. The dark signal is directly related to the exposure time, which the user is able to manipulate. For cooled cameras operated with moderate exposure times (<1 s), the dark signal does not significantly contribute to the temporal noise[15,50]—rather, its contribution is to spatial noise, which we discuss in Sec. II B 2.

### 2. Spatial noise

Ideally, all pixels in a given array would behave identically under identical conditions—this, unfortunately, is not the case. Spatial noise refers to the discrepancy between pixels, which may introduce unwanted non-physical patterns in captured data. In contrast to temporal noise, which fluctuates around the genuine signal, spatial noise cannot be "averaged out" in the same way. Spatial noise is especially relevant to low-light imaging as its influence relative to the signal in this regime makes it increasingly apparent,[51,52] and structured noise can rapidly degrade the visual quality of an image.

One approach to analyzing different sources of spatial noise is by use of "noise maps" on the camera, which can then be used to calculate "non-uniformities," which characterize that source of spatial noise.[51,52] The different noise maps considered include dark noise, read noise, and photoresponse. Details on how to gather, generate, and characterize noise maps can be found in Appendix A 4.

Dark signal non-uniformity (DSNU) refers to discrepancies between the offset values of pixels. While, in principle, the offset is merely an artifact of the digitization procedure, the offset value can still differ from pixel to pixel for a number of reasons. One being the thermal dark signal—these are electrons that become excited not due to incident light but due to thermionic excitation.[51]

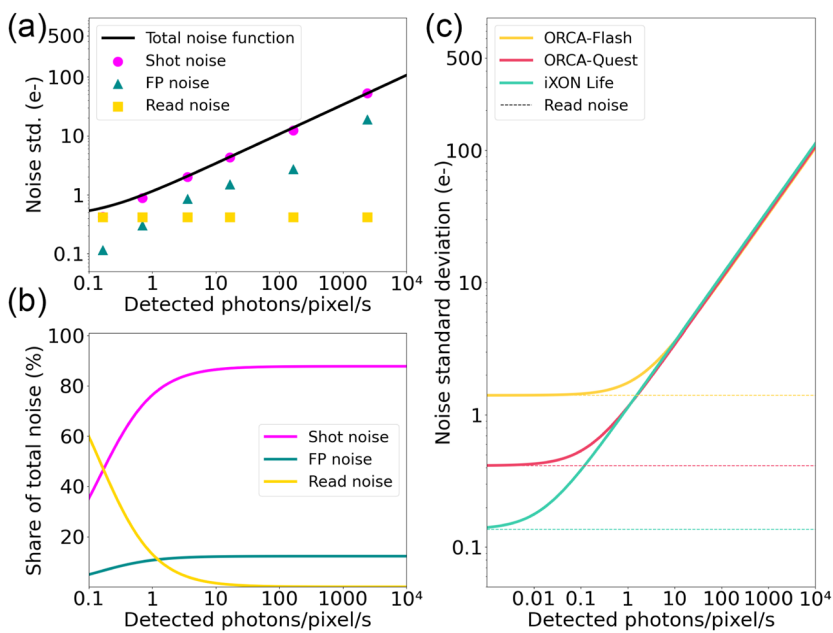

**FIG. 5.** Total noise function and its use in assessing noise sources and comparing different cameras. (a) The total noise function for the ORCA-Quest, with the empirically determined standard deviation of the different noise components plotted against it. This is referred to as a photon transfer curve. (b) Break down of noise predominance as a function of incident intensity for the ORCA-Quest. It should be noted that shot noise dominates outside all but the extreme low-light region. (c) Total noise curves for the three cameras considered in this tutorial. The effect of read noise on the noise floor is particularly apparent with this comparison.

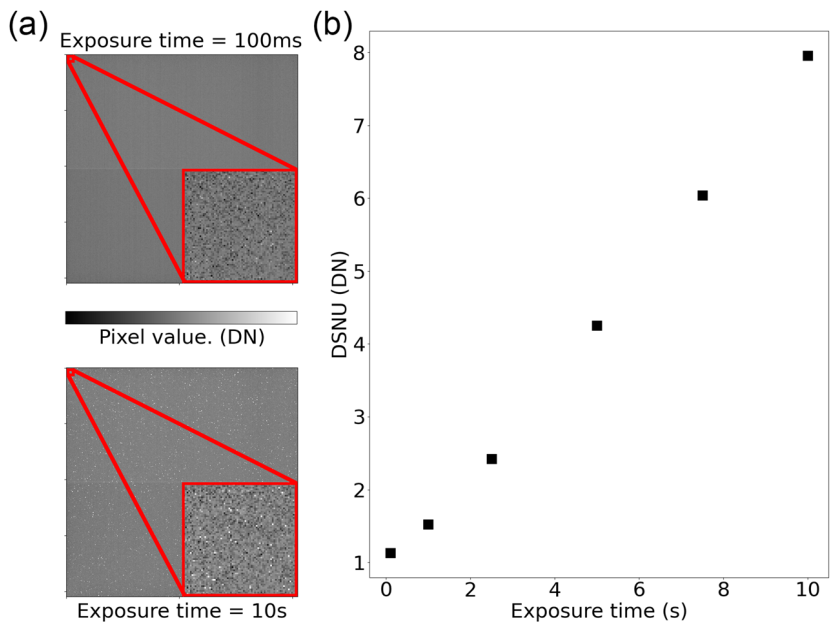

FIG. 6. Dark signal/offset spatial noise, derived by averaging entirely dark frames. (a) Offset maps captured with the ORCA-Flash at 100 ms and 10 s exposure time. At 100 ms, there is spatial noise present, but primarily in fixed rows and columns; at 10 s, there is a greater addition of noisy pixels. See the insets for an expanded view of the pixel values. (b) DSNU as a function of exposure time. The value is almost eight times higher at 10 s of exposure time when compared to 100 ms of exposure time.

Figure 6 contains noise maps from the ORCA-Flash; this camera is air cooled to −10 °C, which is a higher temperature than the ORCA-Quest at −20 °C, and the iXON Life, which can be cooled down to −80 °C. Both temperature and exposure time affect how much dark noise will be present and thus affect a given captured image. For quantitative microscopy, it is advisable to create an "offset map" for your camera, which will allow you to subtract the per-pixel offset of your digital image.

Photoresponse non-uniformity (PRNU) characterizes the difference in value reported by pixels illuminated by the same amount of light. It is often reported as a percentage, with 100% referring to the mean pixel response; a photoresponse of 103% would suggest that a given pixel exaggerates the observed intensity by 3%. Figure 7 compares the photoresponse noise maps for an sCMOS camera to that of an EMCCD camera. We can see that, despite the absence of temporal fixed-pattern noise in CCD sensors, patterned artifacts can be present due to non-uniformity in the photon response introduced during the annealing stage of sensor fabrication.[53] In fact, this camera exhibits a higher PRNU than the sCMOS cameras. Confining our attention to Fig. 7(a), we can see some camera artifacts are due to contaminants on the glass shield in front of the sensor; this is after extensive cleaning of the camera and is primarily caused by solvent residue and lens cloth fibers. This demonstrates that physical contaminants can have a relevant effect on imaging uniformity.

The pixels in Fig. 7 do not demonstrate more than a few percentage points of difference between them. In the context of low-light imaging, dealing with a limited number of photons per pixel, the effect of this form of spatial noise is negligible.

Although "read noise" is a form of temporal noise, the read noise map and non-uniformity (RNNU) identify pixels that are more (or less) noisy than is typically the case. One major benefit is that it allows the identification of "hot pixels": these refer to pixels which are dramatically noisier than is typical for a given platform. These hot pixels are disproportionately responsible for camera noise, and many sCMOS cameras are equipped with automatic hot pixel correction, which attempts to combat this issue. Figure 8 depicts how these noise maps can be used to identify noisy pixels. Figure 8(c) in particular depicts the distribution of noisy pixels for the ORCA-Flash.

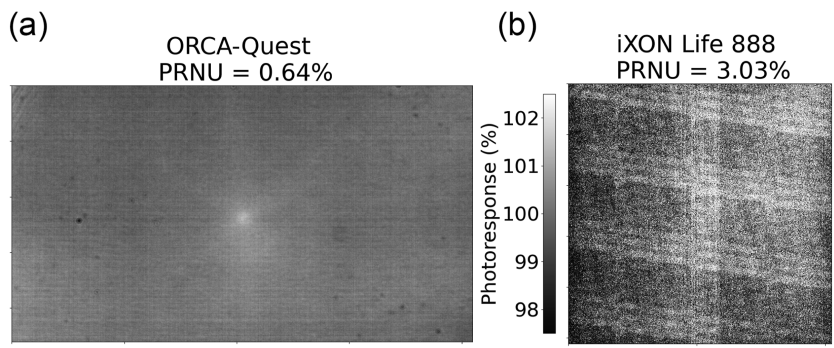

FIG. 7. Photoresponse spatial noise. (a) Photoresponse map for the ORCA-Quest. The PRNU for the camera is low (<1%), so low that spots of difficult to remove residue, and reflections from internal metal casings are visible. (b) The photoresponse map of the iXON Life 888 is more distinctive, with the structured pattern a consequence of the fabrication of the sensor.

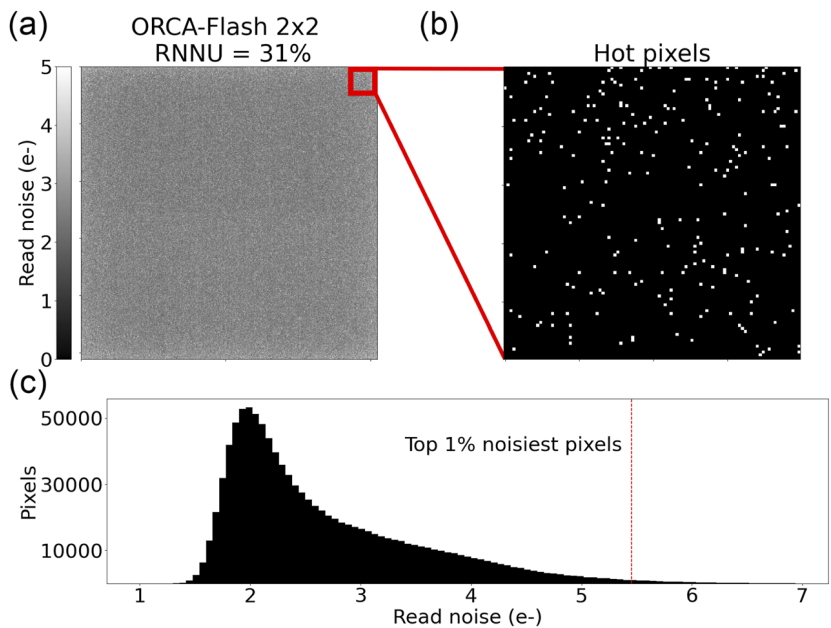

**FIG. 8.** Non-uniform read noise spatial noise. (a) Read noise map for the ORCA-Flash in 2 × 2 pixel binning mode, exhibiting a spread in the read noise per pixel. (b) The top 1% noisiest pixels from the highlighted region of the noise map, referred to as "hot pixels." (c) Histogram of read noise distribution; the mean value for the ORCA-Flash in 2 × 2 pixel binning mode is 2.8 e−, but with a "fat tail" distribution.

Although spatial noise is relevant to low-light imaging, modern scientific camera arrays are highly uniform, and the effect of spatial noise should not be exaggerated; even data-intensive quantitative imaging modalities such as super resolution microscopy are often more susceptible to the negative effects of temporal noise rather than spatial.[52] One major exception is the presence of "hot pixels," which occasionally report values dramatically higher than is physically meaningful. If not properly dealt with, these may invalidate entire datasets based on a single outlier. Spatial noise also has a complex relationship relating to post-processing denoising algorithms, which will be discussed in Sec. III.

## III. FLUORESCENCE IMAGING

While Sec. II A is dedicated to the theoretical understanding of camera operation and noise, this section demonstrates the effects in a practical imaging scenario. A purpose-built two-photon LSFM as described in Appendix B 1, and illustrated in Fig. 9, is used to capture data. LSFM is an increasingly popular modality due to its low photodose and inherent optical sectioning abilities, making it especially suited for imaging of live samples.[10,12] Key to this is the use of wide-field fluorescence imaging, which in combination with plane illumination leads to rapid volumetric imaging.[10]

In the following experiments, live blastocyst-stage embryos were illuminated by a 740 nm ultrashort pulsed source, exciting autofluoresence primarily from NAD(P)H and FAD;[7,21] details of the preparation can be found in Appendix B 2. Although some points of discussion apply exclusively to the light sheet geometry (namely, confocal line scanning/"light sheet mode"), most conclusions are applicable to wide-field fluorescence microscopy more generally.

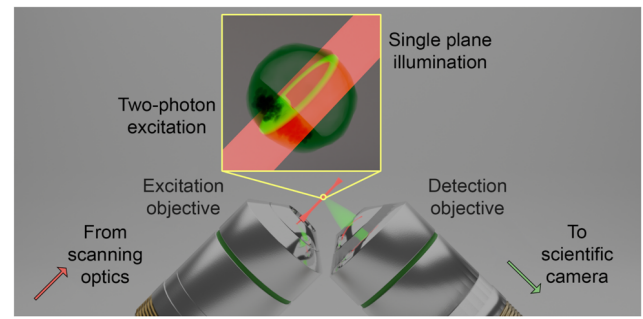

**FIG. 9.** Geometry of scanning light sheet fluorescence microscopy. Perpendicular objectives allow for the excitation of the fluorophores and subsequent capture of their emission from only a single plane using a "light sheet." The light sheet is formed by scanning a beam (using a galvo mirror) parallel to the detection plane of the objective at high-speed relative to the camera exposure. The resulting fluorescence is then captured by using a scientific camera. Moving the sample through the light sheet allows us to capture planes sequentially to generate a 3D image.

Two-photon LSFM utilizes near-infrared pulsed excitation to achieve increased penetration and has been used to improve imaging at depth in live samples.[54–59] Two-photon microscopy, particularly in light sheet mode, is especially suited for autofluorescence imaging due to its improved gentleness and penetration,[10,44] while light sheet has demonstrated low photodamage to live samples.[12] However, the lower two photon absorption cross section[44,60] of many fluorophores leads to a lower fluorescence signal, which can be a limiting factor in quantitative measurements. Overall, this platform is well-positioned to be representative of low-light wide-field microscopy, where the imaging platform would benefit from careful camera choice to capture low intensity signals for more efficient imaging. Thus, this is the approach that we use for imaging in this tutorial.

To characterize imaging quality, we employed the live mouse embryo as our sample. All mammals begin as a single cell, formed by the fusion of an oocyte and a spermatozoon at fertilization. This one-cell embryo then undergoes a series of cell divisions until a large fluid filled cavity forms approximately 4–5 days post-fertilization (species dependent). At this stage, the embryo is termed the blastocyst and is comprised of distinct cell lineages: the inner cell mass and trophectoderm. It is at this stage of development that an embryologist chooses a single embryo—from a cohort of embryos—to transfer into the patient in an IVF cycle. The recent surge in non-invasive, optical approaches to diagnose embryo viability[4,6,46,61,62] together with the sensitivity of the early embryo to light[12,63] make this an ideal test case for exemplifying our camera architectures. In particular, capturing autofluorescence (label-free) has emerged as a promising route for embryo analysis as it can capture signals from NAD(P)H and FAD, among others, which link directly to metabolic activity and viability. Thus, here we use the blastocyst-stage murine embryo as an exemplar for our imaging studies of low-level light (autofluorescence) to elucidate the camera principles and optimize image capture.

All the images are captured under a fixed magnification of 40× and cropped for presentation. Pixel size is a major determinant of imaging quality, as will be discussed in Sec. III C, and for that reason, results are alternatively reported in units of photo-electrons/pixel/s or photo-electrons/$\mu m^2$/s depending on which measure is more relevant. Due to continued development of the live samples, imaging of embryos occurred within an eight-hour window of development to limit variability in structure and fluorescent emission. It should be noted that photoelectron values are obtained by multiplying a pixel's average DN value by the conversion factor $K_{ADC}$, which produce fractional results. While the number of photo-electrons is quantized, and therefore, always an integer, we are considering the average over a number of frames, which is not necessarily a discrete value.

## A. Image quality assessment

Assessing image quality is a non-trivial task, with several different metrics being developed, each possessing different definitions and use cases.[64,65] Generally, these metrics seek to characterize the relative predominance of noise effects within the image, but they can differ subtly. It is a popular choice to characterize image quality using the peak signal-to-noise (PSNR),[52,64,66] a log-scale metric in units of decibels (dB) based on the ratio between the brightest pixel values and the mean-squared error of the noise. When assessing cameras, it is more typical to use "signal-to-noise" (SNR),[16,17,25,40] which is the ratio between the signal and its noise standard deviation. There also exist more sophisticated metrics, such as the Structural Similarity Index (SSIM), which characterizes the similarity between two images (in this case, a ground truth and noisy image) for luminance, contrast, and structure[65]—it is common to see this metric used along with others such as PSNR,[52,66] and it has been demonstrated they share an analytical relationship.[67]

In order to best characterize imaging quality for quantitative microscopy, two metrics are selected: signal-to-noise ratio (SNR), and the closely related contrast-to-noise ratio (CNR), expressed as follows:

$$\text{SNR} = \frac{S}{\sigma}, \quad \text{CNR} = \frac{S - B}{\sigma}. \quad (9)$$

Here, $S$ is the "signal" of the image, $B$ is the value of the background, and $\sigma$ is the noise standard deviation, consistent with the noise model of Sec. II B 1.

There are a number of reasons which recommend this approach: SNR has a more straightforward physical significance, where an SNR < 1 implies a signal below a temporal noise floor, while the PSNR, relying on the "peak" of brightest pixels, is susceptible to "hot pixel" events, as described in Sec. II B 1. Furthermore, the SNR is consistent with the total noise function [Eq. (8) Sec. II B 1]; this model of noise is "agnostic" about the origin of the detected light—the CNR allows for accounting of a "noise background" present in the image. Not only does the noise background degrade visual quality but also may confound quantitative microscopy by adding an additional "offset" to ratiometric analysis. The CNR provides a measure of both visibility and also reliability for quantitative microscopy.

The equations presented in 9 are standard definitions for these metrics,[64,68] but there is considerable subtlety in rigorously defining these terms. As such, a procedure for empirically determining these metrics was devised, partially drawing on the previous work of *Joubert and Sharma*.[40]

We begin the process by capturing a stack of 100 images under identical conditions, as seen in Fig. 10(a). In order to fairly define SNR independent of camera-specific parameters such as pixel size or gain, rather than work with digital numbers, we convert the images within our stack into units of normalized pixel intensity,

$$I_{i,F} = \frac{\text{DN}_{i,F}}{g \times A_{px} \times t_{\text{exp}}} \times 64 \mu m^2, \quad (10)$$

where $I_{i,F}$ is the value of the $i$th pixel in the $F$th frame in units of photo-electrons/s/pixel, for a pixel of 64 $\mu m^2$ in area. This area was selected because it is approximately the average pixel size of the cameras considered in this tutorial. As we can assume that each image captures the same underlying field-of-view, with differences caused by noise, we can define the signal and noise as the average and standard deviation of the pixel across the stack, respectively,

$$\bar{I}_i = \text{Avg}_F[I_{i,F}], \quad \sigma_i = \text{Std}_F[I_{i,F}]. \quad (11)$$

This can be seen in Fig. 10(b), showing the signal and noise determined per pixel using this method.

One issue with defining SNR for fluorescent samples is the high-contrast nature of fluorescence microscopy—bright "signal" pixels contrast against dark pixels. If we were to include these "dark pixels" within the analysis, then the overall SNR would be considerably lower. Therefore, it is necessary to select only the "signal" pixels, which are above the noise background.

In order to do this, a "dark" region of the image stack is selected, as in the top image of Fig. 10(c), and the background intensity is defined as

$$I_B = \text{Avg}[I_{i,F}], \quad i \in \text{Background region}. \quad (12)$$

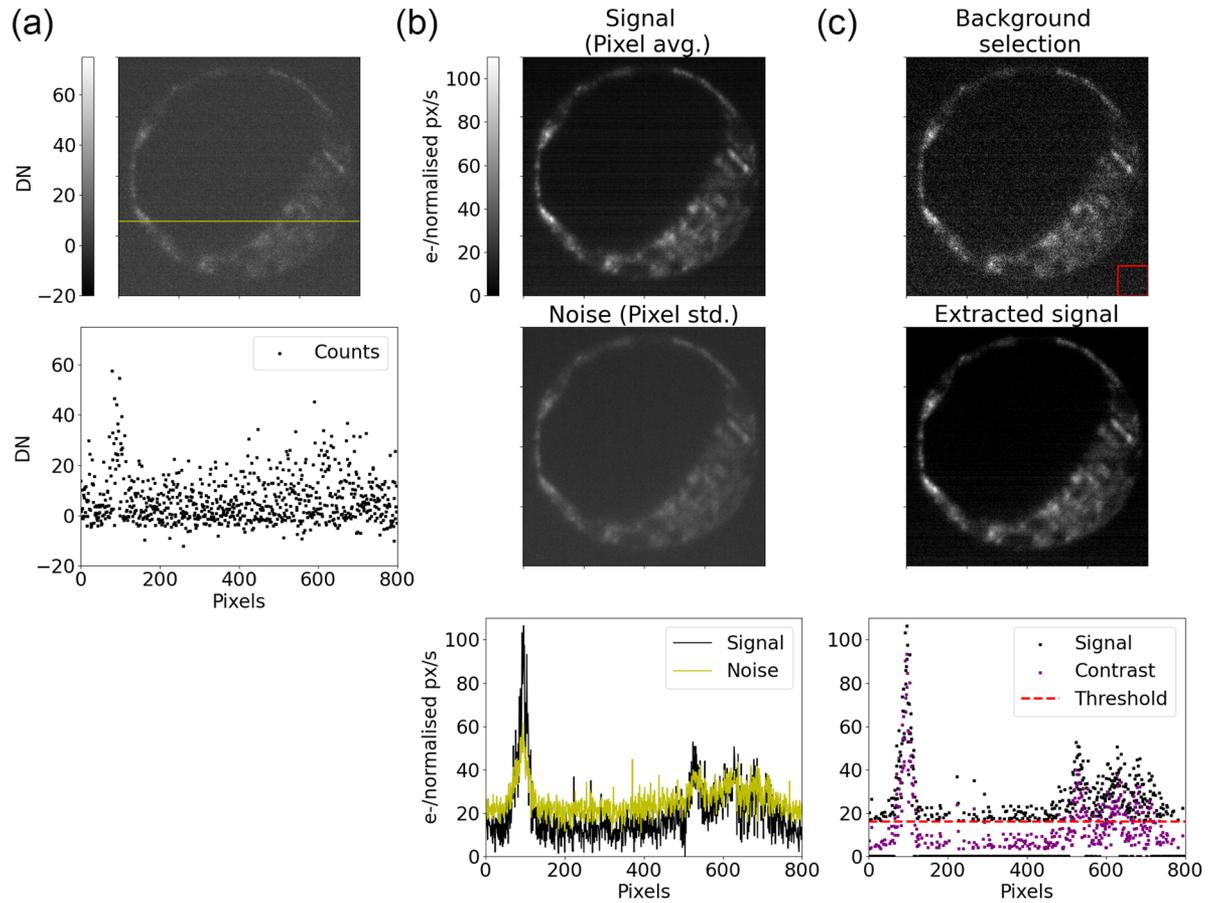

**FIG. 10.** Steps taken to determine the signal-to-noise and contrast-to-noise ratios for a stack of images captured under identical conditions. In order to limit the effect of sample mobility to the temporal noise, acquisition times were limited to a maximum of 30 s. (a) An image of a mouse blastocyst-stage embryo captured with a light sheet microscope; the trace in the following shows the digital number value for each pixel within the highlighted row. This is one image in a stack of identical images. (b) Defining the signal and noise based on the pixel average and standard deviation of the sample stack. The pixel values have been converted to normalized pixel intensity, and are plotted together in the following. (c) The determination of the threshold based on the background value. $I_B$ is given by the average normalized pixel intensity in the highlighted region [see Eq. (12)], while the threshold is equal to $I_B + \sqrt{I_B}$; the signal extracted is those pixels with signal values over the threshold. To determine the contrast, $I_B$ is subtracted from the signal image $I_i^S$, as can be seen from the trace in the following.

Assuming the background noise follows a Poisson distribution, we can select for the pixels one standard deviation above the background intensity by setting the threshold as

$$I_i^S = \bar{I}_i > I_B + \sqrt{I_B}. \quad (13)$$

With the signal per pixel defined as such, the SNR and CNR for the images within the stack can be calculated as

$$\text{SNR} = \text{Avg}_i\left[\frac{I_i^S}{\sigma_i}\right], \quad (14)$$

$$\text{CNR} = \text{Avg}_i\left[\frac{I_i^S - I_B}{\sigma_i}\right]. \quad (15)$$

The mean signal intensity per unit area can then be calculated as

$$\bar{I}^S = \text{Avg}_i\left[\frac{I_i^S}{64\ \mu m^2}\right]. \quad (16)$$

Equations (14)–(16) provide empirical measurements, which we are able to compare against the predictions of our noise model,

$$\text{Predicted SNR}(\bar{I}^S) = \frac{\bar{I}^S \times A_{px} \times t_{\exp}}{\sigma_{e-, \text{total}}(\bar{I}^S \times A_{px} \times t_{\exp})}. \quad (17)$$

CNR can be predicted in a similar manner; while the background intensity contributes to the noise, it does not contribute to the contrast. We can, therefore, adjust the predicted SNR by a factor given by

$$\text{Predicted CNR} = \frac{\bar{I}^S - I_B}{\bar{I}^S} \times \text{Predicted SNR}, \quad (18)$$

where $\bar{I}^S$ and $I_B$ are in equivalent units. This coefficient characterizes the portion of the signal which is not from the background.

A closely related metric to SNR is the *relative* SNR, or rSNR, which is defined as

$$\text{rSNR} = \frac{\text{SNR}}{\text{iSNR}}, \quad (19)$$

where "iSNR" is the ideal SNR, which is free of all camera noise and background, with shot noise being the only contribution. Therefore, we can formulate it as

$$\text{iSNR}(\bar{I}^S) = \frac{\bar{I}^S \times A_{px} \times t_{\exp}}{\sqrt{\bar{I}^S \times A_{px} \times t_{\exp} \times g^2}}. \quad (20)$$

With Eqs. (17), (19), and (20), we can derive the following expression:

$$\text{rSNR}(\bar{I}^S) = \frac{\sqrt{\bar{I}^S \times A_{px} \times t_{\exp} \times g^2}}{\sigma_{\text{total}}(\bar{I}^S \times A_{px} \times t_{\exp})}. \quad (21)$$

Here, rSNR characterizes the portion of image noise which can be attributed to the camera, with 1 denoting perfect performance. We observe that for the "relative CNR," the ideal CNR is equal to the ideal SNR (no camera noise or background), allowing us to define

$$\text{rCNR} = \frac{\bar{I}^S - I_B}{\bar{I}^S} \times \text{rSNR}, \quad (22)$$

which includes the effects of noise background on camera noise.

To reiterate, this procedure allows for the empirical calculation of SNR and CNR for fluorescent images within a stack [Eqs. (14) and (15)]. These empirical results can be compared against the predictions based on the camera noise model [Eqs. (17) and (18)], which can then be used to determine the portion of noise attributable to the camera itself [Eqs. (21) and (22)]. This procedure for the empirical determination of SNR, CNR, and average sample power has been adapted as a Python notebook, which can be accessed in the supplementary material, along with some sample data for testing.

It must be noted that this method is susceptible to perturbations of the underlying sample, such as the movement of live samples; motion causes change in the underlying pixel, the value of which will be interpreted as noise. For this reason, each acquisition was limited to a maximum of 30 s, a time span over which the sample we use is effectively immobile.

While SNR and CNR are useful metrics for quantitative microscopy, they are only partially predictive of "visual quality."[65,66] Spatial noise especially can degrade the visual quality while minimally affecting either, which has motivated the development of other metrics such as the SSIM, given by the following formula:[65]

$$\text{SSIM} = \frac{(2\mu_r \mu_t)(2\sigma_{r,t})}{(\mu_r^2 + \mu_t^2)(\sigma_r^2 + \sigma_t^2)}, \quad (23)$$

where $r$ and $t$ are the "reference" and "test" images, respectively, linearized into a one-dimensional array. $\mu$ is the mean of either array, while $\sigma$ is the standard deviation, and $\sigma_{r,t}$ is the covariance between the two.

However, the SSIM relies on the use of a reference image: in this case, a ground truth without any spatial noise, which we do not have access to. Despite this, SSIM still presents a useful metric for the use of denoising algorithms, to be discussed in Sec. III G. Due to the propensity for "hallucinations" of detail from deep learning approaches, the SSIM provides a method of comparing structural content in denoised images.

### B. Comparison of cameras

In order to characterize the imaging quality of scientific cameras, manufacturers typically include figures of merit, such as the read noise and QE. While these parameters are indeed meaningful, they are not necessarily immediately predictive of actual performance. The method of quality assessment discussed in Sec. III A enables a method of fair comparison between different cameras. As described above, we captured autofluorescence from live mouse embryos using LSFM for the different cameras under analysis. Details for the preparation of these samples can be found in Appendix B.

A first analysis of Fig. 11(a) would suggest that the iXON Life far outperforms the two sCMOS cameras for samples of the same intensity, which is consistent for both the empirically measured and predicted SNR. However, when we consider not the absolute intensity, but intensity relative to pixel size as in Fig. 11(b), we see that iXON Life's superior imaging performance is not due to its camera architecture, but rather due to its larger pixels (169 $\mu m^2$ compared to 36 and 21 $\mu m^2$ for the ORCA-Quest and ORCA-Flash, respectively). Under this analysis, it can be seen how imaging performance converges on the basis of intensity per pixel despite the considerable differences between the cameras. This highlights how important pixel size is for imaging quality, which will be expanded upon in Sec. III C.

The relative SNR/CNR plots allow us further insight into the imaging quality; from Fig. 11(c), it is seen that in the absence of any noise background, the iXON Life and ORCA-Quest exhibit comparable rSNRs within this intensity regime, implying they should perform similarly for the same pixel size. However, the excess noise of EMCCD cameras, $F_n$, inflates the effect of background noise, reducing its rCNR relative to sCMOS cameras. This excess noise term is independent on the level of gain used—one, therefore, cannot just turn the electron-multiplying gain down to reduce the excess noise while retaining the increased signal.

Figure 12 provides a useful comparison between an sCMOS and EMCCD camera, as with 2 × 2 pixel binning (see Sec. III C) the two cameras share identical pixel sizes. In this figure and throughout the tutorial, we make use of "noise traces" to assist in visualization of both the signal and noise for a cross section of an image. The plot displays the signal of the pixels, with the thickness of the line reflecting the standard deviation about the mean.

Considering the noise traces of Figs. 12(a) and 12(b), it can be seen that the pixels in Fig. 12(a) exhibit higher variability, i.e., noise despite the signal being higher relative to the read-noise floor. The closeness to the noise floor can be noticed through close inspection of the image in Fig. 12(b), which reveals structured noise in the background, due to its sCMOS architecture. These results are consistent with the calculated SNR/CNR and the relative SNR/CNR plots of Fig. 12(c)—the iXON Life exhibits a higher SNR and rSNR, but the effect of excess noise decreases the rCNR relative to the ORCA-Flash. However, it can also be seen that the ORCA-Flash, nearer its read noise floor, would exhibit a more precipitous decline in performance were the signal power to decrease.

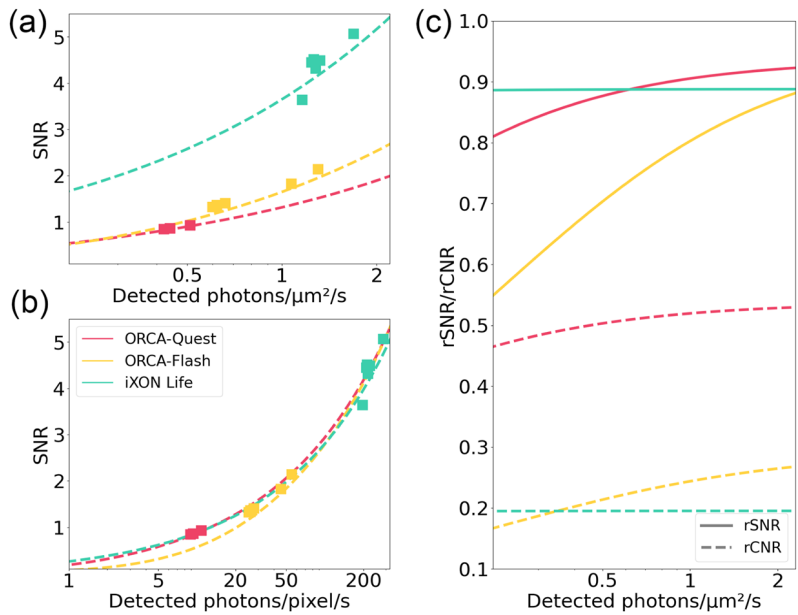

**FIG. 11.** Empirically determined image SNR for cameras under analysis, using $1 \times 1$ pixel binning and an exposure time of 100 ms. The background coefficient for the rCNR curved is based on the average background coefficient of the images for that camera. (a) Measured SNR for several images plotted against the predicted SNR curve for the respective cameras as a function of sample intensity per unit area. (b) The same comparison, plotted as a function of intensity incident per pixel. (c) The relative SNR and CNR curves of the cameras for this intensity range. The CNR shows the effect of a fluorescent background on the imaging quality that we may obtain.

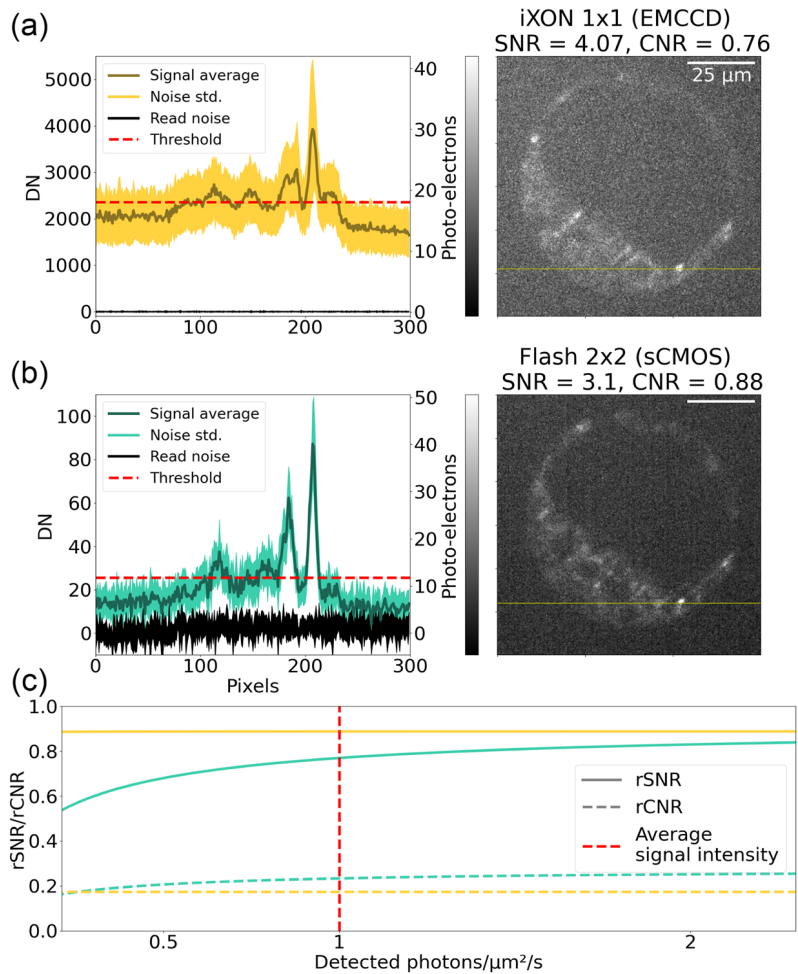

**FIG. 12.** Noise trace comparison of a mouse blastocyst-stage embryo image captured with an EMCCD and sCMOS cameras with equivalent pixel size, based on 100 captured frames. The yellow line on the image denotes which pixel row is being depicted in the trace. The thickness of the trace is equivalent to the noise standard deviation for the pixel within the image stack, while the central, darker line indicates the average value. Exposure time for both the images is 100 ms. (a) Image captured by iXON Life, along with its noise trace; due to the EM gain, the signal is substantially higher than the noise floor. (b) The same embryo imaged with the ORCA-Flash in $2 \times 2$ pixel binning mode. The signal is much closer to the noise floor, but the features of the trace are more distinct due to the absence of excess noise. (c) Relative SNR and CNR curves for the two cameras with the sample intensity indicated. The excess noise from the EMCCD camera inflates the noise background, leading to a lower CNR.

It is clear from this that while read noise and QE are key predictors of quality, other factors such as pixel size or excess noise are just as significant when it comes to practical imaging. Section IV will discuss how key figures of merit should be considered when selecting cameras.

### C. Pixel size, pixel binning, and magnification

As highlighted in this tutorial, pixel size is a major determinant of quality when capturing an image. For this reason, manufacturers frequently include the ability to "bin" pixels, or combine multiple pixels on the pixel array, effectively increasing the pixel area. This can be a powerful resource for improving image quality. Consider the noise traces from Figs. 13(a) and 13(b), where the four-fold increase in pixel size results in an increased SNR, with the pixel trace reflecting this improvement. The trade-off is a reduction in the image resolution from 900 × 900 pixels to 450 × 450 pixels.

Both sCMOS and CCD cameras are capable of pixel-binning, and different manufacturers allow for different degrees of flexibility in this regard. For example, the sCMOS cameras in this tutorial are limited to 1 × 1, 2 × 2, and 4 × 4 pixel binning, whereas the iXON Life 888 is able to bin any valid tessellating shape, such as binning entire rows. When pixel binning with sCMOS cameras, due to the on-pixel read-out electronics, the pixel read noise is increased; this can be seen in Fig. 13(c), where 2 × 2 binning effectively doubles the read noise standard deviation for sCMOS cameras—CCD cameras, due to their architecture, do not accrue this additional read-noise. This would seem to make CCD cameras more appealing for using pixel-binning—generally, this is true, but typically CCD arrays already possess larger pixels, making binning both less necessary and "coarser," in the sense the binned pixels will be particularly large.

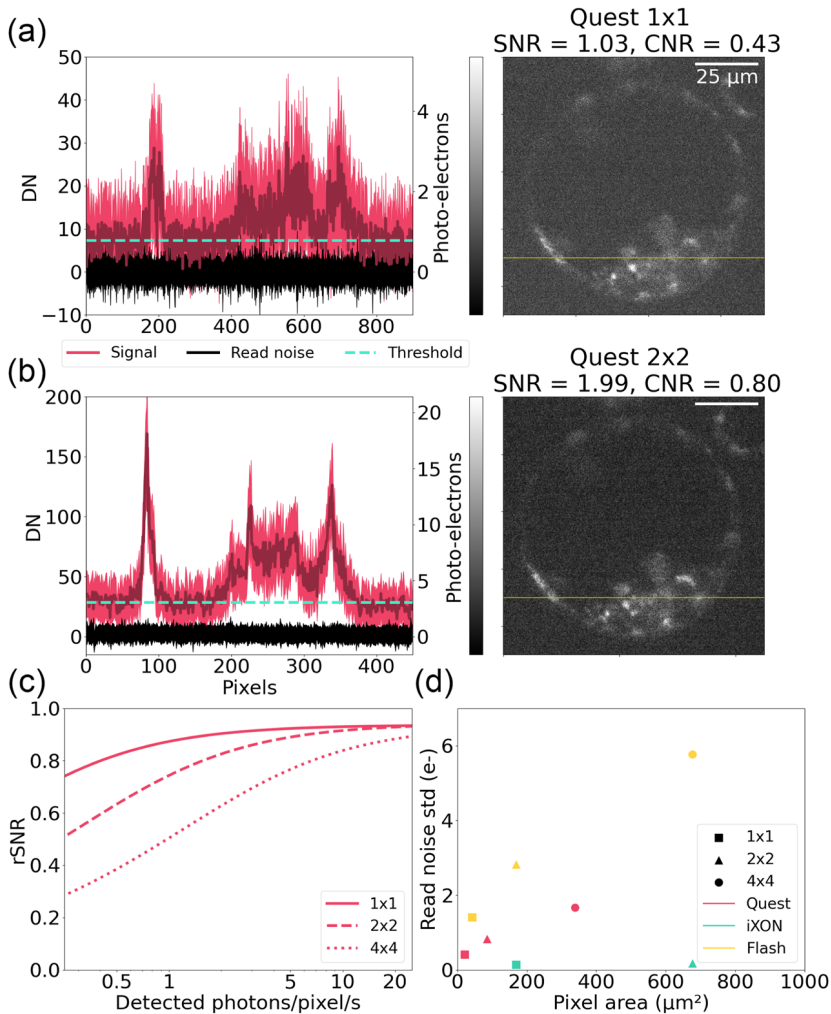

**FIG. 13.** Comparison between ORCA-Quest images captured in 1 × 1 pixel binning and 2 × 2 pixel binning mode. (a) Noise trace and image of an autofluorescent embryo captured with 1 × 1 pixel binning. (b) Noise trace and image for the same embryo, captured with 2 × 2 pixel binning. The increased sharpness of the cellular features in the trace is immediately apparent—the consistency of the threshold between the two traces is also apparent. (c) Increased read noise from pixel binning leads to a reduction in the imaging quality. This can be seen in the reduction of the rSNR curve. (d) The read noise of different cameras as a function of pixel binning—for a sCMOS camera, the read noise increases with the square root of the area, while this effect is not present for a CCD camera. It should be noted that the 4 × 4 data point for the iXON is out of range and not included.

The "trade-off" for pixel binning is that larger pixels result in a coarser sampling of the incident light field. However, it should be noted that this does not necessarily result in a loss of spatial information: Fig. 14(a) demonstrates how the minimum spatial resolution attainable is limited by the diffraction limit; greater magnification beyond the Shannon–Nyquist limit "oversamples" the incident light field, leading to "redundancy" in spatial information. This is known as "oversampling," also called "supersampling." This diffraction limit is determined by the numerical aperture (NA) of the optical system and the wavelength of light within the image and is typically within the range 0.4–1 $\mu$m. In cases of oversampling, pixel binning, despite coarser sampling, does not necessarily reduce the spatial information present in the image. This is most relevant for optical systems with low numerical aperture and high magnification.

In this tutorial, considering the numerical aperture of the system (see Appendix B 1), along with the approximate wavelength of the fluorophores (500 nm), the diffraction limit can be estimated at 310 nm. Given the magnification of 40×, this is magnified to a diameter of 12.4 $\mu$m; Appendix A 1 contains details on camera pixel and sensor dimensions, and in their 1 × 1 pixel operating modes, the iXON is undersampled and the ORCA-Flash is close to optimal sampling, while the ORCA-Quest images are oversampled. For that reason, there is only a 33% resolution reduction rather than 50% seen in Fig. 13(b).

As opposed to pixel-binning (increase in pixel size at slightly increased read-noise), when possible, the reduction of magnification is a similarly effective method of improving SNR; Fig. 14(b) shows how smaller pixels are able to capture an identical resolution with a reduced magnification. Therefore, smaller pixels need not inherently result in decreased SNR—this discussion will be continued in Sec. IV.

Finally, one thing to note is the possibility of "pixel crosstalk," present in highly sensitive sCMOS cameras, where charge accumulated on one pixel is read-out on adjacent ones; this effect potentially limits spatial resolution, especially when the scene is undersampled.[69] This crosstalk effect is less predominant in the photon-sparse regime, as there are fewer charges to be transposed across pixels.

Pixel binning is an effective method of increasing the image SNR by sampling more light at the expense of reduced resolution and (for sCMOS cameras) increased read noise. This is especially attractive for cases where "oversampling" is occurring; the pixel size and degree of sampling are typically overlooked in discussions of optimizing camera performance, but are a significant factor, especially in the low-light regime.

### D. Exposure time

Changing the exposure time is the most obvious approach to controlling the collection of light from a scene. The longer a pixel is active, the more photons are able to interact with it and more signal is able to be collected. This can be seen in Figs. 15(a) and 15(b), where increasing the exposure time by a factor of three results in a higher SNR. It is also apparent that the increased light collection has also included a substantial background portion, meaning the CNR has not increased to the same degree. Figure 15(c) shows how increased signal collection results in a better rSNR deeper into the low-light regime, but the rCNR does not see a similar increase.

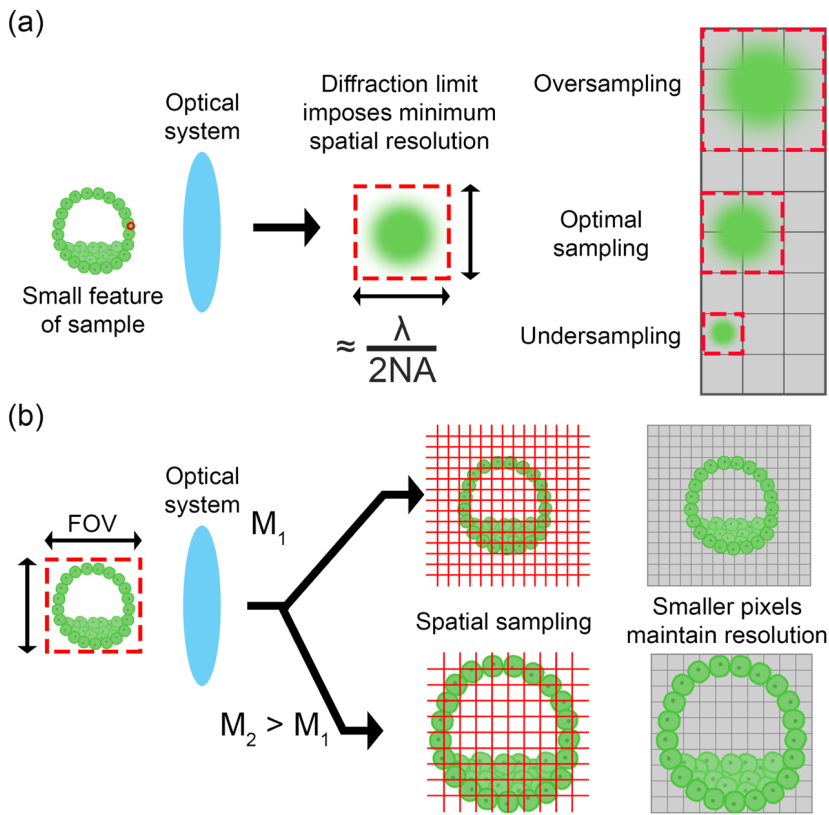

**FIG. 14.** Relationship between image resolution, magnification, and spatial resolution. (a) The physical nature of optical imaging places a limit on the spatial resolution that can be captured by the system, referred to as the diffraction limit. The Shannon–Nyquist criterion states that a diffraction-limited spot should be sampled twice in both spatial dimensions; greater magnification leads to oversampling, which while increasing the image resolution is not able to resolve any further spatial detail. Pixels that are too large may result in undersampling, in which case, there is a loss of spatial resolution. Here, we see what happens when we image the small region encircled in red on the blastocyst and the corresponding image on the camera for the three cases: undersampling, optimal sampling, and oversampling. (b) Smaller pixels are able to resolve identical spatial resolutions for the same FOV at a reduced magnification.

Increasing the exposure time is generally the preferred method of increasing signal capture; as discussed in Sec. II B 2, longer exposure times increase the chance of registering dark signal events within the frame—while for exposure times <1 s, this effect is insignificant, and even for longer exposure times, the increased signal is often worth the trade-off. However, increased exposure times obviously have an effect on the camera frame rate and may be unsuitable for samples that are in some way mobile or dynamic, such as live samples. One of the major appeals of LFSM is its use in volumetric imaging, where large image stacks are collected, each frame accumulating for the set exposure time; these lengthy timescales can be an issue, especially for live samples, where photodamage and photobleaching is a concern. These cases may call for other approaches to optimization such as pixel binning.

Finally, one further aspect to consider is the camera full well depth. Using long exposure times can lead to saturated pixels even when capturing a low-intensity scene. This is especially a concern when a high background is present, and the signal may not be able to "out-compete" the background noise.

### E. Operating modes

Modern scientific cameras are increasingly equipped with alternative "operating modes," which alter some aspect of camera operation to optimize imaging for a particular scenario. In this section, we will confine our analysis to the different operating modes represented by the ORCA-Quest: the standard mode, also called the "area mode," along with its unique "photon resolving" mode, and sCMOS-exclusive "light sheet" mode. Other camera platforms may offer analogous modes of operation.

"Photon resolving" mode for the ORCA-Quest reduces the camera read noise to such a level that the user is able to distinguish between one vs multiple photons incident on a given pixel. This mode is exclusive to qCMOS cameras,[49] but similar "noise reduction" or "quiet scan" operating modes may be present on other scientific cameras. This mode imposes a restriction on the exposure time, limiting it to a minimum of 11.52 ms per row within the imaging region. Here, its most interesting feature is its reduction of read noise, and this will be explored further in Sec. III F.

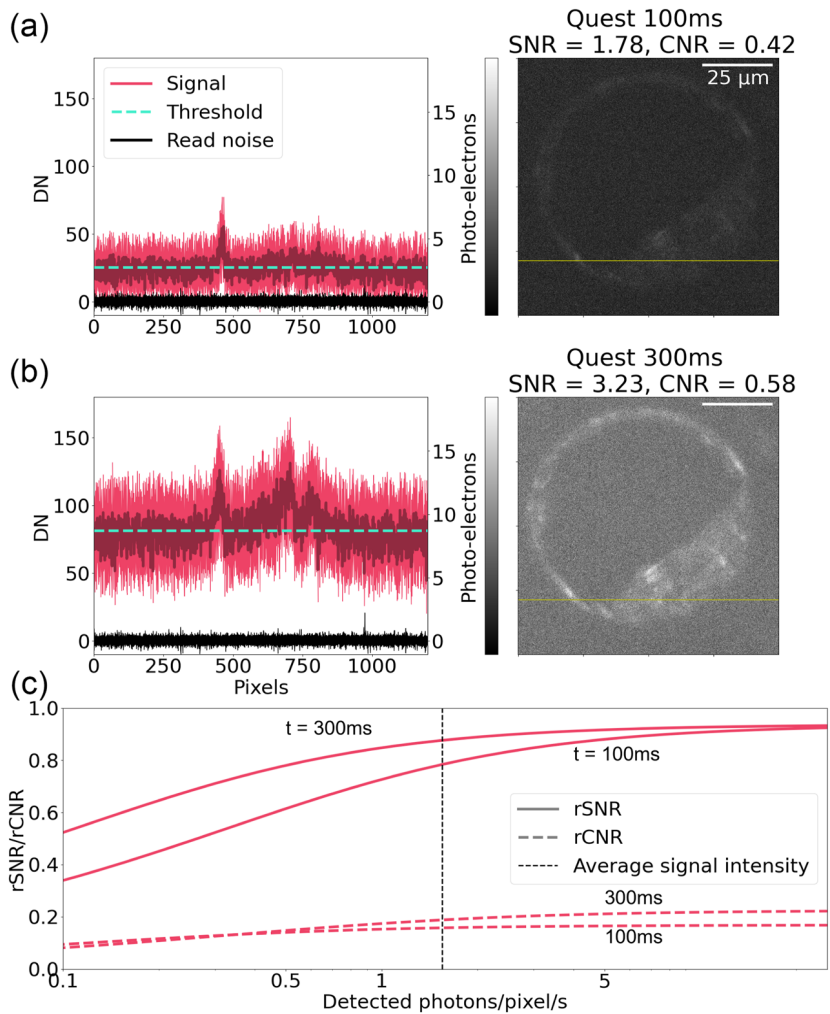

**FIG. 15.** Comparison between images of a mouse blastocyst-stage embryo captured with two different exposure times. (a) Image and noise trace of an autofluorescent blastocyst-stage embryo imaged with an exposure time of 100 ms. (b) The same embryo imaged under identical conditions, with an exposure time of 300 ms. Longer exposure time allows for more signal intensity to be captured, while also capturing more background signal. (c) Relative SNR and CNR curves for the ORCA-Quest for differing exposure times. Due to the increased collection of background noise, the improvement for the CNR is less than that for the SNR.

"Light-sheet mode" is another mode provided by certain sCMOS cameras, including the ORCA-Quest and ORCA-Flash, which enables user manipulation of the row read-out speed for the rolling shutter. This allows for the read-out of pixel rows to be synchronized to the scanning of the virtual light sheet; this is depicted in Fig. 16(a) and is referred to as "confocal line-scanning."[26]

The advantages of this synchronization mode can be seen in Fig. 16(b): in its regular operating mode, the light sheet is generated by rapidly scanning the laser over the pixel array; for a given exposure time, the beam will illuminate the same region multiple times. The scan rate must be higher than the exposure time of the camera to ensure the entire plane is illuminated and sampling artifacts are avoided. This means that scattered fluorescence from the sample can also fall incident on the pixel row while active, contributing to background noise.[26] Light sheet mode allows the exposure time to be decreased without reducing the incident exposure from the relevant sections of the sample. Due to the demands of temporal synchronization, the period of the light sheet must be equal to the frame period; for the ORCA-Quest, the read-out time can be set as slow as 275 ms for a full frame read-out, which would demand a light sheet scan rate of only 3.6 Hz. This is in comparison with the much faster scan rates used for regular operating modes, where the light sheet scan rate must be much greater than the camera frame rate.

Figure 17 contains a comparison of the same autofluorescent sample imaged using different camera operating modes of the ORCA-Quest. This is a particularly dim sample, below the temporal noise floor, where a switch to a quieter mode may be desired. Figure 17(a) represents the standard mode of the camera and should, therefore, be the primary point of comparison. Figure 17(b) depicts the "photon resolving" mode of the camera, and in this mode, the digital pixel numbers are equivalent to generated photo-electrons. It should be noted that the read noise floor has lowered, which reduces the noise in the pixel row trace, and subsequently improves the SNR, although only by a very modest amount; this mode is effective via its reduction of read noise, which leaves the shot noise contribution. Although this mode is primarily aimed at photon counting and quantum imaging cases, the reduced noise is beneficial to imaging generally.

Figure 17(c) depicts light sheet mode, which successfully rejects much of the background illumination. This can be seen in the noise trace, which looks "noisier" as without the Poissonian background signal, it is now composed primarily of read noise; this can be seen in the structured background noise. Although the signal is lower for this case, the primary reduction in intensity is due to the rejection of scattered noise, leading to the highest CNR of any image captured. This confocal line-scan approach offers a promising method for the almost total rejection of scattered background noise in LSFM, but the limitations imposed on pixel binning and exposure time may need to be compensated in other ways, such as reduced magnification. Light sheet mode is also offered on the ORCA-Flash camera, which is a popular camera for use in LSFM setups and similar modes may be found on other camera platforms.[10] Although the sample considered is especially dim, the reduction in read noise or background provided by the operating modes applies exactly as it does for brighter samples.

### F. Optimization of camera performance

The previous sections have sought to illuminate the different imaging parameters that a user has at their disposal to improve image quality. Figure 18 depicts how the selection of the correctly "optimized" settings can achieve this, without recourse to increased illumination.

Figure 18(a) depicts an image of a sample captured with unoptimized camera settings; Fig. 18(b) is captured using the same settings, except the embryo is illuminated by approximately double the excitation intensity. While there is a visually apparent improvement in quality, along with an increased SNR/CNR, the higher intensity leads to an increased background, so the improvement in CNR, for instance, is minimal. Furthermore, these improvements come at the cost of increased photobleaching and toxicity.

Compare this to Fig. 18(c), where the power is the same as in Fig. 18(a), except that different imaging parameters were used. The longer exposure times allow for greater collection of light, placing the signal above the read noise floor (which is already reduced by the camera mode) and the 2 × 2 binning still provides enough spatial resolution to image small features.

Finally, Fig. 18(d) compares the position along the SNR/CNR curves of the various images. While increasing the power shifts the position along the curves (while also affecting the CNR due to increased background), the optimized settings are able to achieve superior SNR and CNR despite lower power.

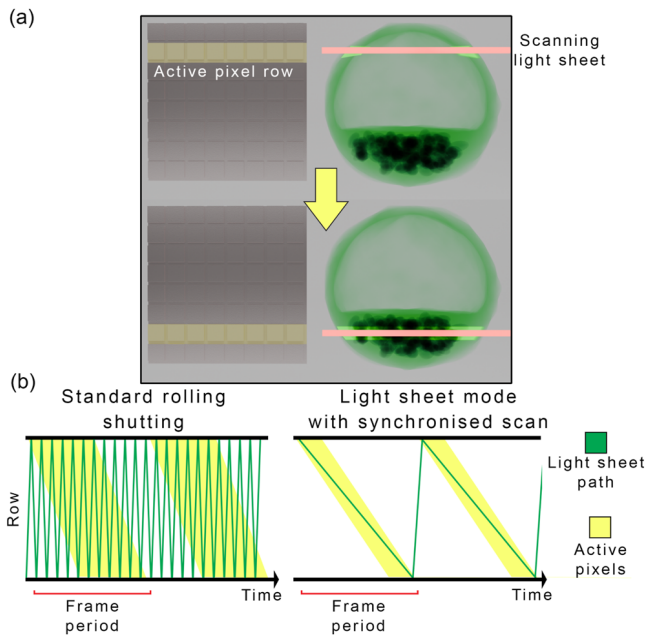

**FIG. 16.** Light sheet operation mode, also known as confocal line scanning. (a) Diagram of how light-sheet mode enables the synchronization of the virtual light-sheet scanning with the pixel row read-out of a sCMOS camera. Each row observes only one pass of the beam as it is exposing, limiting the collection of scattered photons. (b) A timing diagram depicting synchronized confocal line-scanning. The path of a virtual light-sheet in standard operating mode is compared to its path when synchronized for light-sheet mode; rather than capturing the emission from multiple passes, only a single pass occurs during the entire frame period.

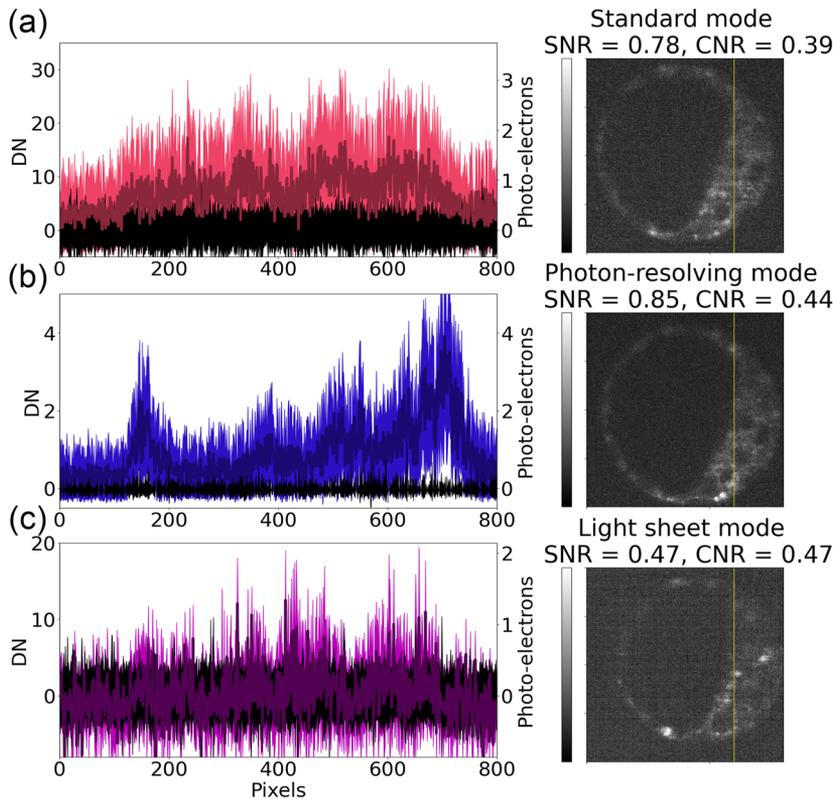

**FIG. 17.** Comparison of the mouse blastocyst-stage embryo imaged across different operating modes of the same camera. (a) Noise trace and image recorded for the ORCA-Quest in standard mode. Exposure time = 100 ms (b) The same sample imaged with the camera in the low-noise "photon resolving" mode. Exposure time = 104 ms (c) The same sample now imaged using the "light sheet" operating mode, with synchronized light-sheet scanning. Exposure time = 3 ms.

In order to fairly compare the images in Fig. 18, it is necessary to perform a contrast adjustment of the images—that is, to change the maximum and minimum grayscale values to provide maximum contrast for the features of interest. This is performed in Fig. 19, where the histograms for the images are displayed. It should be noted that due to the operating mode and pixel binning, the histogram for the optimized settings is remarkably different despite capturing the same fluorescence. When imaging, it is necessary to keep in mind that the apparent visual quality is in large part determined by the contrast, and adjusting the contrast is necessary to fairly compare disparate images.

This section has demonstrated how the correct selection of imaging settings can be used to improve the imaging quality without recourse to increased illumination, which can lead to photobleaching and photodamage.

### G. Post-processing of image data

Recent years have seen dramatic growth in the popularity and sophistication of machine-learning algorithms, especially in "machine vision" and image recognition.[70] This has driven interest in "denoising" algorithms, which seek to identify noise artifacts within an image in order to suppress them. While a full discussion of denoising biological images is beyond the scope of this tutorial (see Laine et al.[66]), it is germane to discuss these denoising algorithms in the context of low-light fluorescence imaging, armed with our knowledge as above.

In order to investigate the effectiveness of denoising algorithms in relation to different sources of camera noise, a small selection of algorithms was tested on images captured with an sCMOS camera in both its regular and light sheet operating modes, as well as an EMCCD camera. This covers cases when the background noise is entirely Poissonian (EMCCD), entirely structured ("light sheet mode"), and a mixture of the two (sCMOS). Each dataset consists of 100 images captured under identical conditions, consistent with the other datasets analyzed here.

Three diverse, representative algorithms were selected: Noise2Fast,[71] Neighbor2Neighbour,[72] and Accurate Correction of sCMOS Noise (ACsN[50]). Noise2Fast is noted particularly for its speed, while Neighbor2Neighbor makes use of a unique sampling method, and ACsN is uniquely suited for sCMOS cameras. These are all "non-supervised" algorithms, meaning that the training set does not contain any clean, ground truth data. In contrast to the other two, the ACsN algorithm also takes into account specific camera noise via a calibration procedure. Data collected from the photon transfer curve was used to create the gain and offset maps necessary for the algorithm. Details on how these algorithms were applied can be found in Appendix C.

Each image is denoised independently, meaning the training set consists solely of an individual image, creating a stack of 100 denoised images. Due to the different denoising procedures used, the digital number values are not conserved; the denoised stack is normalized and rescaled toward the same value range as the original images. The SSIM reported is the average SSIM for each image within the stack taken with respect to the uncorrected signal image, composed of the average of the uncorrected image stack.

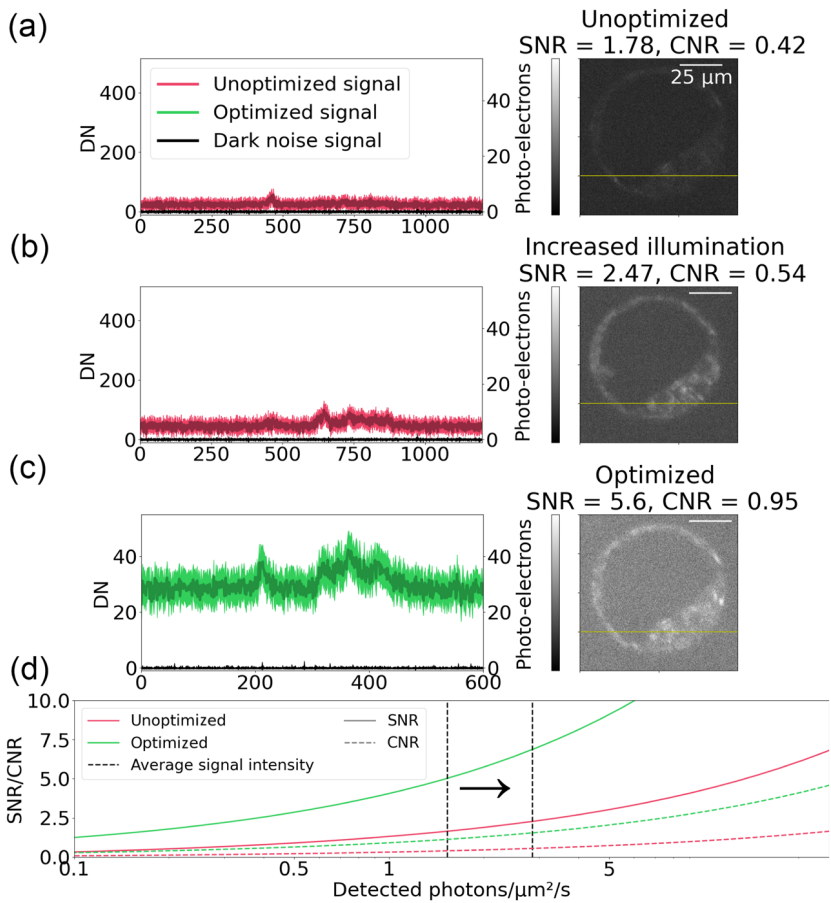

FIG. 18. Comparison between the effects of increasing illumination intensity and altering camera imaging parameters. The ORCA-Quest was used for these measurements. (a) Embryo imaged with $1 \times 1$ pixel binning and 100 ms exposure time in the standard operation mode. (b) The same embryo imaged under identical conditions, with approximately double illumination intensity. (c) Embryo, imaged in $2 \times 2$ pixel binning, 250 ms exposure time, with the photon-resolving operation mode. (d) The SNR and CNR curves for the unoptimized and optimized camera settings. The arrow depicts the shift of the average signal intensity due to the increased illumination.

Figure 20 presents a comparison of the different algorithms on different test cases. Considering for a moment the ACsN algorithm, we see that for the two images captured using the ORCA-Quest, there is a marked increase in SNR and CNR, whereas for the iXON Life, the effect is negligible. This is to be expected, as ACsN is explicitly tailored for the reduction of sCMOS noise and in fact lends confidence to the operation of the algorithm. It can be seen that the SSIM is unchanged in the case of the standard mode, or slightly reduced in the case of LS mode; for the latter, this decrease is likely from the removal of structured noise in the background. These facts point toward ACsN being an effective method of denoising sCMOS images—when combined with confocal line-scanning with light-sheet mode, it offers a potentially very powerful resource for high quality images at low-light levels.

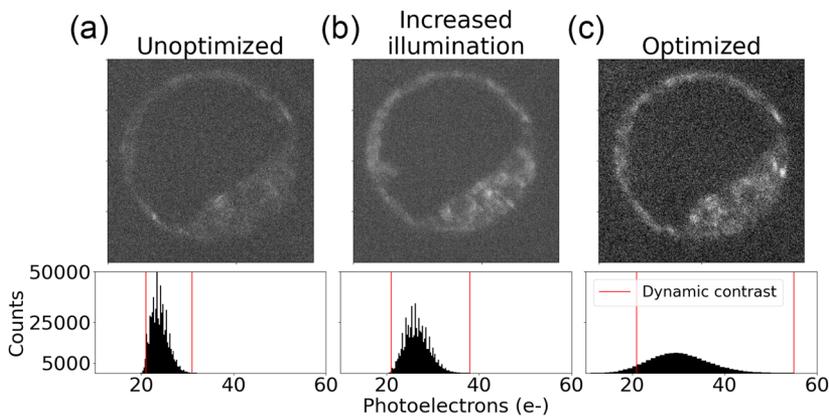

FIG. 19. Comparison of contrast-adjusted images from Fig. 18, with (a), (b), and (c) corresponding to the images of that figure. Beneath each image is a histogram of their pixel values in units of photo-electrons with the dynamic contrast highlighted—the minimum and maximum grayscale values in the image are determined by the range of the dynamic contrast. As individual images only contain the relative intensity information, dynamic contrast allows optimal contrast for features under analysis.

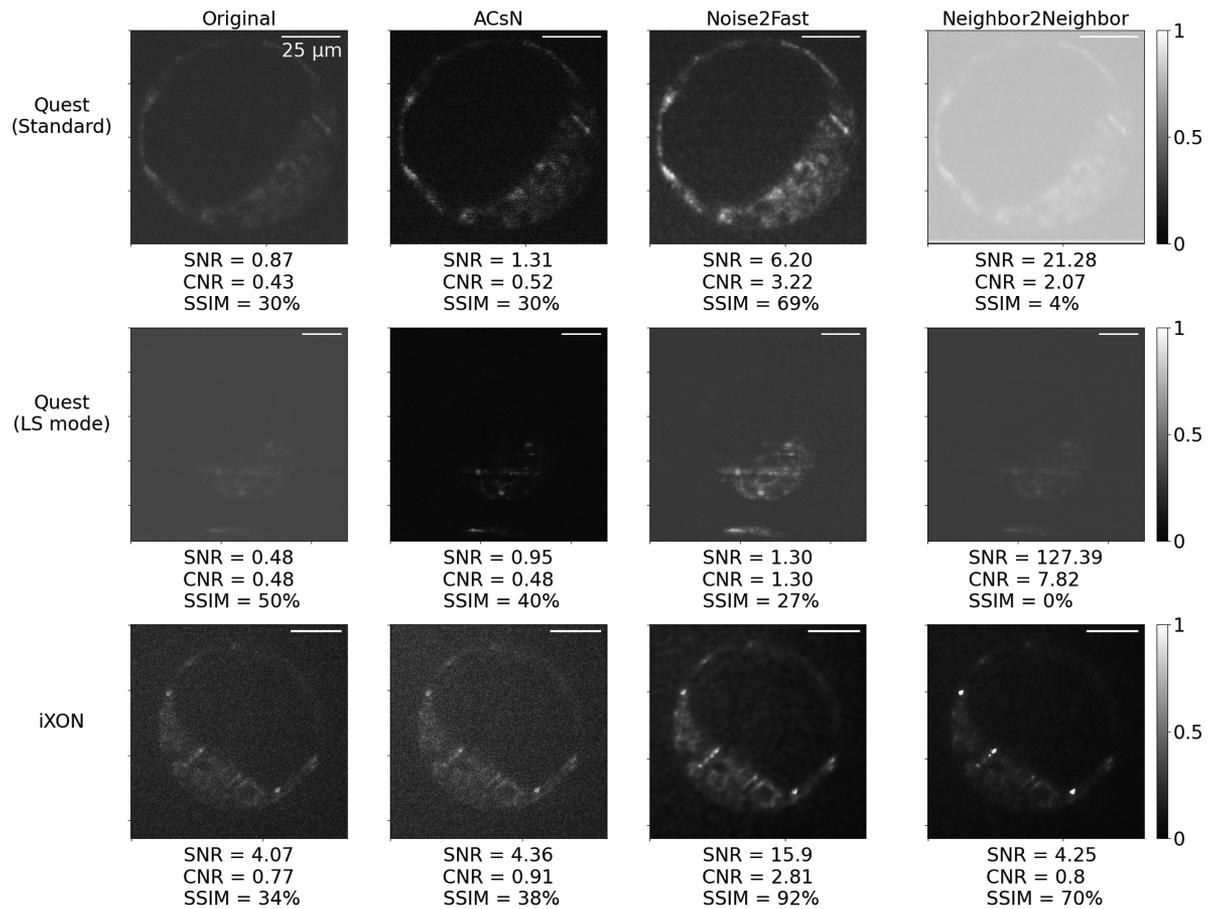

**FIG. 20.** Effect of the three different unsupervized denoising algorithms discussed in the main text on images of mouse blastocyst-stage embryos captured under differing conditions. The ORCA-Quest camera was run in "Standard" and "light sheet" (LS) modes vs the iXON camera. SNR, CNR, and SSIM all provide different insights into the effect of the algorithm; the high SNR resulting from the Neighbor2Neighbor algorithm being applied to the light-sheet mode image suggests high performance, but the low SSIM implies that the extracted signal is not of that embryo.

In contrast to ACsN, Noise2Fast and Neighbor2Neighbor are generic unsupervised denoising algorithms, which apply no knowledge of camera physics. Rather, deep learning is applied in an attempt to capture the underlying structure of the image in order to determine genuine features from noise. From Fig. 20, the Noise2Fast algorithm appears to successfully increase the image quality across most metrics for all cases considered, and visually there is an increased contrast for the embryos against the background. In comparison, Neighbor2Neighbor presents more complex results; while for the iXON image, it offers a moderate improvement, the sCMOS images appear "washed-out" with reduced contrast, despite dramatically high SNR values. The SSIM values are illustrative here: compared to the ORCA-Quest (Standard) results, we see there is an increase in SSIM, as well as SNR—this suggests that the denoising was successful, more closely resembling the signal image. In the Neighbor2Neighbor case, this is not true, implying that the increase in SNR is due to a poor reconstruction of the underlying structure. The case for ORCA-Quest (LS) is a bit more complex, as although there is an increase in SNR while using the Noise2Fast algorithm, the SSIM has decreased relative to the original or even the ACsN algorithm; while this may be due to a reduction in the structured background, it could also suggest that "overfitting" has occurred, where there is a reduction in temporal noise, but the structure does not reflect that of the embryo. Nevertheless, the performance is superior to Neighbor2Neighbor, which has significantly reduced SSIM and exaggerated SNR. For the iXON, the Noise2Fast algorithm exhibits very strong performance, with all metrics dramatically improved, while the Neighbor2Neighbor algorithm exhibits only moderate improvements.

The dramatically differing performance of the algorithms can be explained by the different methods of subsampling. While Noise2Fast uses "checker board" downsampling, Neighbor2Neighbor, as its name suggests, uses "neighbor" downsampling, where the image is divided into a grid, with neighboring pixels being selected from it. This random element in neighbor downsampling may be negatively affected by the presence of sCMOS spatial noise; as this noise manifests in row- and column-based patterns, and by confining our selection to neighboring pixels, we introduce structured noise components into the training set. This explains why the reconstruction for the ORCA-Quest (LS) image is so poor, whereas for the iXON image, the effect is minimal. This suggests that the architecture of the camera used has implications for which algorithms will be effective for denoising, with the method of downsampling being especially relevant when dealing with the case of structured noise.

Of the samples denoised in Fig. III G, those considered for the ORCA-Quest have low SNR; this makes them especially challenging for the denoising algorithm. Theoretically, brighter samples with higher SNR have more apparent structural details, which makes identifying noise features easier—despite this, the relative performance of the Noise2Fast algorithm is superior for the ORCA-Quest ($\approx 7\times$ SNR improvement) than for the iXON ($\approx 4\times$ SNR improvement). While one factor may be the excess noise of the EMCCD camera, another is reduced resolution of the iXON image, both in terms of spatial undersampling and reduced number of pixels. This means there is less structural data for the algorithm to be trained on, suggesting a complex relationship between resolution, SNR, and denoising performance; a full analysis of which is outside the scope of this tutorial.

The status of denoising algorithms is complex and evolving rapidly. Visual analysis and the use of quality metrics lend support to their effectiveness, but ultimately, the aim of imaging is to reconstruct a luminous intensity field, and no algorithm is able to provide more insight into the underlying physical reality than is already present. "Hallucinations," the generation of false features, is an ever-present concern—one exacerbated by the presence of structured, patterned noise of the kind present in low-light imaging with sCMOS cameras.[66] Even an algorithm such as ACsN, which utilizes an understanding of camera physics still relies on an inherently statistical process which may introduce errors. There has been application of denoising algorithms for the improved extraction of functional metrics in quantitative microscopy;[73] such approaches must be treated very carefully, as the post-processing algorithm has no access to the underlying physical reality, which is the concern of quantitative analysis. "Denoising" is not a substitute for genuine reduction in noise, and using deep learning to extract parameters that are not directly found within an image is methodologically troublesome.

## IV. DISCUSSION AND CONCLUSION

This tutorial has sought to familiarize the reader with key concepts in digital imaging, including the nature and origin of camera noise, and provide them with tools to achieve high quality imaging under low-light conditions for wide-field microscopy. The case used to demonstrate the various points was two-photon light sheet microscopy, which is a popular, powerful fluorescence imaging approach. Image optimization depends heavily on use case: cameras are only able to record what is incident on their sensor, the origin of which they are entirely agnostic about. For this reason, it is necessary to have a clear idea of what is sought to be captured in an image before optimization can take place.

Before imaging, let alone optimization, a camera must be selected. In order to communicate the imaging power of their cameras, several figures of merit are used, including the QE, read noise, resolution, and pixel size. For most modern scientific cameras, the QE across the visible spectrum is typically quite high; the difference in detection efficiency offered by a camera with 95% QE as opposed to 80% is minor and could easily be compensated with a slight increase in exposure time. Rather, a larger concern is for quantitative microscopy with multiple spectral components, where QE curves may confound ratiometric analysis.

Similarly, read noise is a significant figure of merit, which must be placed in context. All else being equal, the read noise will determine the noise floor of the camera, but this value is defined only for a single pixel. Low read noise may be necessary for photon counting experiments, but fluorescence microscopy differs from such studies, as the number of photons collected is of no consequence—rather, the objective is to use the light intensity as a means of obtaining spatial information from a scene. As such, decreased read noise may not be an improvement if the collection efficiency is reduced.

This leaves the resolution and pixel size as figures of merit, which we have discussed above in Sec. III C and will be discussed further in the following. The selection of cameras is a subtle task, and QE and read noise are both key figures for merit for judging their imaging power. However, for low-light fluorescence microscopy, small differences in QE or read noise by themselves will offer only minor improvements.

Another example of the subtlety in camera selection is the distinction between sCMOS (or qCMOS) and EMCCD cameras; while the latter attain a level of sensitivity unreachable for sCMOS cameras, the excess noise may diminish the reliability of these results, especially for samples with a noise background. This was seen in Fig. 12, where the signal captured by the sCMOS camera was much closer to the camera noise floor, but due to amplification of the noise background did not have a higher CNR.

As discussed in Sec. III A, the "background coefficient" used to determine the CNR is defined as $B = \frac{\bar{I}^S - I_B}{\bar{I}^S}$. If we assume the excess noise applies to the background intensity and not the signal, we can quantify the effect of the excess noise on this by defining the background predominance as

$$\text{Background predominance} = \frac{\bar{I}^S - F_n I_B}{\bar{I}^S} = 1 + F_n(B - 1). \quad (24)$$

This gives a useful way of comparing EMCCD and sCMOS cameras when exposed to a scene with the same background coefficient $B$. If we take the ORCA-Quest and iXON as representative examples of their respective camera architecture, we are able to compare their performance as a function of incident intensity and background predominance, as can be seen in Fig. 21.

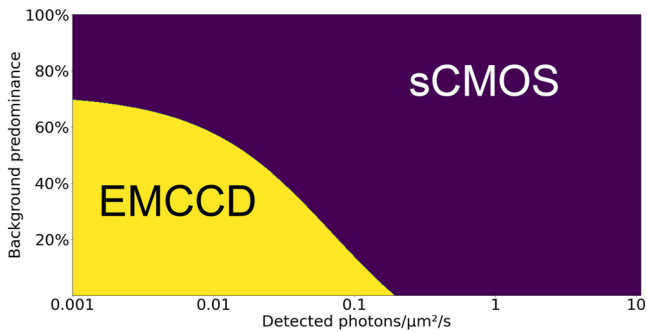

**FIG. 21.** Chart displaying which camera architecture produces a higher CNR as a function of incident intensity and background predominance. Identical imaging conditions are considered, with the ORCA-Quest and iXON Life 888 considered as prototypical cameras of their architecture.

Based on this comparison, EMCCD cameras are effective for dim samples without a predominant background. While EMCCD begins to be the superior choice in the extreme sparse photon regime (<0.1 photons/$\mu m^2$/s), the excess noise makes it so that especially noisy samples (>50% background predominance) become more challenging to capture. However, while the excess noise makes imaging more susceptible to noise background, this background is largely random and Poissonian, in contrast to the structured noise that is present in sCMOS cameras. This structured noise can visually degrade the image quality and is especially confounding for certain denoising algorithms, as discussed in Sec. III G. For applications that are highly sensitive to spatial noise, EMCCD cameras may be preferable.

It is exactly for thick, highly scattering samples where the confocal line-scan "light sheet mode" of certain sCMOS cameras becomes appealing, with its rejection of scattered light via synchronization. However, this mode puts limitations on the pixel size and exposure time and is only compatible with scanned light sheet microscopes. Nevertheless, for obtaining high-quality structural information from fluorescent samples, this technique offers the potential for capture at relatively high speeds, low-light levels, and with minimal photodose.

As previously emphasized, pixel size is a major predictor of image quality, one which is commonly overlooked. "Oversampling," where the diffraction limited resolution of the FOV is sampled above the Shannon–Nyquist criterion, is typically not an issue for microscopy generally, but for low-light quantitative microscopy, the reduction in signal intensity per pixel for no increased spatial resolution is difficult to justify. Figure 22 depicts the level of magnification necessary to capture a high resolution (0.25–1 MPX) scene depending on the underlying field of view aimed to be captured. It can be seen that, especially for cameras with small pixels, oversampling can be reached even for moderate magnifications (≈20×), and in these circumstances, a change in magnification or pixel binning is an option of interest. This shows that despite the apparent drawback of reduced collection ability, cameras with smaller pixels may be an asset as they allow for high-resolution information to be captured with lower magnification, ignoring the complications of high magnification optics.

Of course, the option to change magnification is not always present, and many imaging systems have fixed magnifications while allowing for the exchange of cameras. In these cases, cameras with smaller pixels may need to increase their exposure time to compensate for their reduced collection ability.

Figure 23 is illustrative of this fact. A 1 MPx camera with pixels 8×8 $\mu m^2$ in size and exposing for 100 ms is treated as the benchmark, with its position marked out on the chart with a star. Smaller pixels demand longer exposure times, while larger pixels decrease the resolution of the image. This effect is quite pronounced; considering the range of pixel sizes considered in this tutorial, the exposure time necessary to maintain a similar collection level ranges from 30 to 300 ms, thus spanning an order of magnitude.

While pixel size has generally less flexibility than exposure time as a user-controlled parameter, modern scientific cameras come equipped with a plethora of options, which allow for cropping of the sensor view and binning of pixels; combined with appropriate magnification, this gives the user a number of tools to control the spatial sampling of light. Many cameras which are not particularly sensitive may still be able to capture high-quality images, provided the appropriate choice of magnification, pixel size, and exposure time are used, and we encourage the reader to explore this for their own particular imaging platform.

It is by making these optimizations that camera operating modes such as "quiet scan" or "light-sheet mode" can be used effectively. Although these modes may impose additional limitations on the exposure time or pixel binning which must be taken into account, they can be powerful when used correctly.

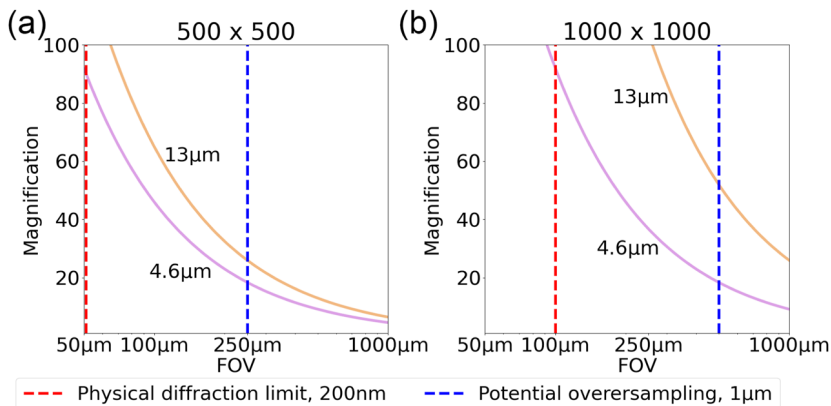

**FIG. 22.** Magnification necessary to achieve an image resolution of (a) 500 × 500 or (b) 1000 × 1000, for cameras of different pixel pitches (4.6 and 13 $\mu m$). The highlighted region between the dotted lines marks where it is possible that the diffraction limit has been reached, depending on the optics used. Once the diffraction limit has been reached, increases in magnification that would produce a higher image resolution will not contain more spatial information.

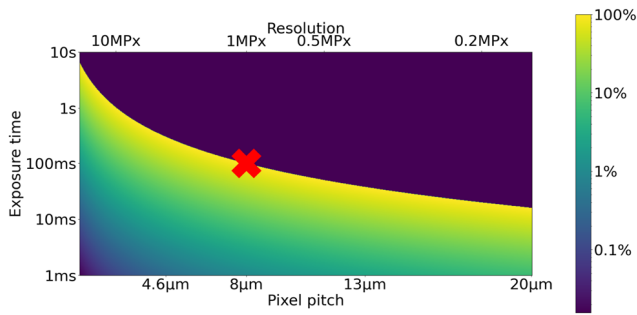

**FIG. 23.** Camera collection efficiency, normalized to a 100 ms image captured with a 1 megapixel (MPx), 8 $\mu$m pixel pitch camera; this location on the plot is marked with a red cross.

Post-processing denoising algorithms, too, may serve as a useful tool in particular circumstances, but it must be emphasized that they should be applied only after all possible imaging optimizations have been performed. The fact that they do not have access to the physical reality of the scene and are able to change pixel digital values in complex, nonlinear ways, makes them methodologically challenging for quantitative microscopy. In scenarios where spatial noise is interfering with visual analysis, they may have useful application, provided the user allows for the possibility of "hallucinations."

Finally, it is necessary to emphasize the importance of understanding camera physics in order to most effectively conduct quantitative microscopy. The digital numbers reported by the cameras require knowledge of the imaging parameters, such as pixel size and exposure time to be physically meaningful, and issues such as pixel saturation, quantum efficiency, and hot pixels are easily overlooked, yet can have major downstream consequences on the data obtained.

## SUPPLEMENTARY MATERIAL

The supplementary material contains a Python notebook with sample data enabling containing code for SNR and CNR calculations.

## ACKNOWLEDGMENTS

The authors thank Paul Wardill and Michael Buckett (Coherent Scientific Ltd.) for their loan of an iXON Life 888 camera. K.D. acknowledges funding from the Australian Research Council (ARC Grant No. FL210100099). K.R.D. and K.D. acknowledge funding from the National Health and Medical Research Council (NHMRC, Grant No. APP2003786). This research was supported by the Australian Research Council Centre of Excellence in Optical Microcombs for Breakthrough Science (Project No. CE230100006) and funded by the Australian Government.

## AUTHOR DECLARATIONS
### Conflict of Interest

The authors have no conflicts to disclose.

### Author Contributions

**Zane Peterkovic**: Conceptualization (equal); Data curation (lead); Formal analysis (lead); Investigation (lead); Methodology (lead); Software (lead); Validation (lead); Visualization (lead); Writing – original draft (lead); Writing – review & editing (equal). **Avinash Upadhya**: Investigation (supporting); Resources (supporting); Writing – review & editing (equal). **Christopher Perrella**: Conceptualization (supporting); Project administration (supporting); Writing – review & editing (equal). **Admir Bajraktarevic**: Investigation (supporting); Resources (supporting); Software (supporting); Writing – review & editing (supporting). **Ramses E. Bautista Gonzalez**: Investigation (supporting); Resources (supporting); Writing – review & editing (supporting). **Megan Lim**: Investigation (supporting); Resources (supporting); Writing – review & editing (supporting). **Kylie R. Dunning**: Resources (supporting); Writing – review & editing (supporting). **Kishan Dholakia**: Conceptualization (equal); Funding acquisition (lead); Project administration (lead); Writing – review & editing (equal).

## DATA AVAILABILITY

The data that support the findings of this study are available from the corresponding author upon reasonable request.

## APPENDIX A: RAW NOISE STATISTICS
### 1. Methods of analysis

Table I contains relevant camera parameters as they appear in provided technical specifications. While this list is not exhaustively descriptive of the cameras, it contains the information most highly relevant for low-light imaging. In order to deepen the characterization of the camera, including for a particular unit, further steps of noise analysis are necessary.

Figure 24 contains a flow chart depicting the method of analysis necessary to determine the various parameters to characterize a camera's noise statistics. While for some parameters, a uniform light source is necessary, for others, it is sufficient to use the camera lens cap to capture dark images. Appendix A 2 will discuss the nature of uniform illumination sources, including their construction; the formulas for calculating the various non-uniformities can be found in Appendix A 4, while the ones for those in the round bubbles can be found in Appendix A 3.

**TABLE I.** Camera parameters provided by technical specifications.

|  | Quest | Flash | iXON (300 gain) |
|---|---|---|---|
| Pixel pitch ($\mu$m) | 4.6 | 6.5 | 13 |
| Max resolution | 4096 × 2304 | 2048 × 2048 | 1024 × 1024 |
| Sensor size (mm) | 18.8 × 10.6 | 13.3 × 13.3 | 13.3 × 13.3 |
| Read noise (e−) | 0.6 | 1.6 | <1 |
| Gain (DN/e−) | 9.35 | 2.17 | . . . |
| QE @ (500 nm) | >80% | >75% | >90% |

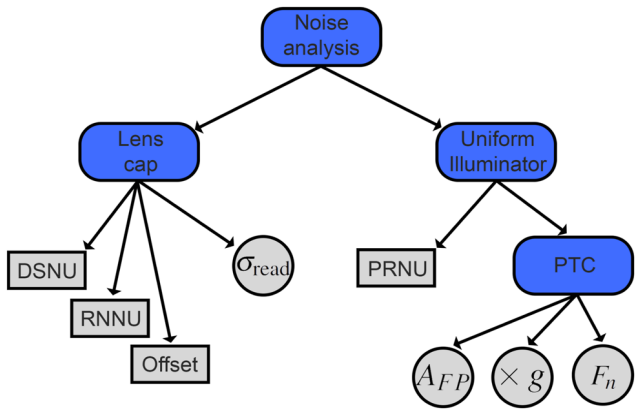

**FIG. 24.** Flow chart depicting how to determine various parameters discussed in Sec. II B.

### 2. Uniform illumination set-up

The objective of a uniform illumination setup is to apply an equal level of intensity to each pixel in the array in order to assess the differences in their response. The primary difficulty is the point-source nature of light sources, which emit spherical, curved wave-fronts that produce intensity gradients when incident on a flat region, such as a camera sensor. This mandates the camera to be positioned at a distance from the source to ensure a flat wave-front, but this comes at the expense of intensity.

Figure 25 depicts a schematic for the uniform illumination setup used for the collection of PTC and noise map data. Two lenses with focal lengths of $f = 50$ mm and $f = 125$ mm are arranged to provide Köhler illumination to the light-shielding box: one before the integrating sphere and one mounted in the input of box. A length of 67 cm allows adequate expansion of the beam such that the light at the aperture is uniform. The light source used is a 527 nm central wavelength LED with a DC power supply, operated at a stable voltage. Neutral density filters at the exit port of the integrating sphere are used to attenuate the input light to appropriate levels.

This source was created with the need to produce highly uniform light across a large dynamic range at different wavelengths of light. Many other methods, including simple application of a mobile phone torch, may be able to produce sufficiently uniform light for purposes of observing contaminants on the glass cover, or even patterns originating from PRNU, but are limited in their dynamic range and introduce confounding factors to their variables.

### 3. Photon transfer curve

#### 1. Procedure

The procedure for collecting PTC is relatively simple: a camera is positioned within the uniform illumination setup, while a combination of ND filters and exposure time is selected to illuminate each pixel to ~90% of the full well depth. A stack of 100 images is then captured, after which another ND filter attenuates the light by an order of magnitude. This process is repeated across five orders of magnitude, which provides the raw data necessary for the analysis covered in the following. This method was adapted from Ref. 15.

#### 2. Analysis

In order to analyze temporal noise, we use the following expression for the DN reported by pixel $i$ for frame $F$:

$$\text{DN}_{i,F} = (S + F_n N_{\text{shot},i,F} + N_{\text{FP},i,F}) \times g + N_{\text{read},i,F} + \text{Off}_i. \quad \text{(A1)}$$

Here, $N$ denotes random variables, which serve as the generator functions for the different sources of noise.

Due to the uniform illumination, we are able to assume that intensity $S$ is independent of the pixel and the frame number.

We find, under no light exposure (such as with the lens cap on), that the digital number is given by

$$\text{DN}^0_{i,F} = N_{\text{read},i,F} + \text{Off}_i. \quad \text{(A2)}$$

Therefore,

$$\text{Off}_i = \text{Avg}_F[\text{DN}^0_{i,F}] \quad \text{(A3)}$$

and

$$\sigma^2_{\text{read},i} = \text{Var}_F[\text{DN}^0_{i,F}]. \quad \text{(A4)}$$

The data available from a single stack of frames, easily captured with the lens cap on, are vital for basic camera statistics. Now, consider a stack of frames captured under a given intensity level $S_n$. The digital number is given by

$$\text{DN}^{S_n}_{i,F} = S_n \times g + F_n N_{\text{shot},i,F} \times g + N_{\text{read},i,F} \times g + \text{Off}_i + N_{\text{FP},i,F}, \quad \text{(A5)}$$

and the total noise at this intensity is defined as

$$\sigma^2_{\text{total},S_n} = \text{Avg}_i[\text{Var}_F[\text{DN}^{S_n}_{i,F}]]. \quad \text{(A6)}$$

In order to isolate the random sources of noise, the difference between subsequent frames is taken,

$$\text{DN}^{S_n}_{i,F} - \text{DN}^{S_n}_{i,F+1} = F_n \times g(N_{\text{shot},i,F} - N_{\text{shot},i,F+1}) + g(N_{\text{FP},i,F} - N_{\text{FP},i,F+1}) + (N_{\text{read},i,F} - N_{\text{read},i,F+1}). \quad \text{(A7)}$$

It can be assumed that due to its origin, the contribution of fixed pattern noise to a given pixel between subsequent frames is negligible,[15] therefore, $(N_{\text{FP},i,F} - N_{\text{FP},i,F+1}) = 0$.

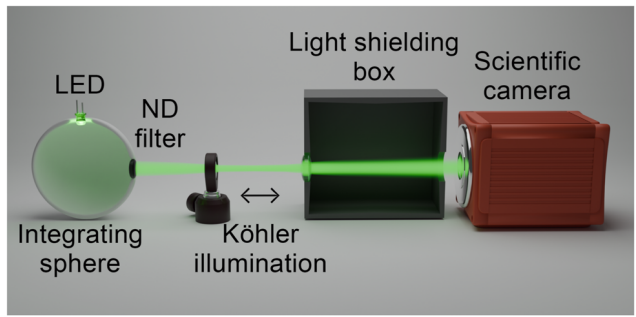

**FIG. 25.** Setup used for the collection of uniform illumination data for the photon transfer curve and noise maps. An integrating sphere with a 527 nm central wavelength LED, and Köhler illumination is used to project a uniform intensity light field into a light-shielding box, after which the cameras under analysis can be placed.

As the difference frames are equal to the difference between two pairs of independent random variables, its variation is equal to

$$\sigma_{\text{diff}}^2(S_n) = 2\big(\sigma_{\text{shot}}^2(S_n) + \sigma_{\text{read}}^2\big). \quad (A8)$$

As a result, we are able to determine the variance of the shot noise as follows:

$$\sigma_{\text{shot}}^2(S_n) = \frac{\text{Avg}_i\big[\text{Var}_F\big[\text{DN}_{i,F}^{S_n} - \text{DN}_{i,F+1}^{S_n}\big]\big]}{2} - \text{Avg}_i\big[\sigma_{\text{read},i}^2\big]. \quad (A9)$$

Finally, we are able to determine $\sigma_{\text{FP},S_n}^2$ as follows:

$$\sigma_{\text{FP}}^2(S_n) = \sigma_{\text{total}}^2(S_n) - \big(\sigma_{\text{shot}}^2(S_n) + \sigma_{\text{read}}^2\big). \quad (A10)$$

Through this analysis, we are able to determine the standard deviations of the different random noise functions at different levels of illumination intensity. Through curve fitting, we are able to determine the parameters discussed in Sec. II B. For example, $F_n$ can be determined by recognizing that

$$\sigma_{\text{shot}}^2(S_n) = F_n S_n, \quad (A11)$$

where $S_n$ can be determined by

$$S_n = \frac{\text{Avg}_i\big[\text{DN}_i^{S_n} - \text{Off}_i\big]}{g}. \quad (A12)$$

Therefore, it is only a matter of performing a linear regression to determine $F_n$. The other variables can be determined similarly,

$$\sigma_{\text{FP}}^2(S_n) = A_{FP} S_n, \quad (A13)$$

and

$$g = \frac{\text{Avg}_i\big[\text{DN}_i^{S_n} - \text{Off}_i\big]}{\sigma_{\text{shot}}^2(S_n)}. \quad (A14)$$

### *3. Parameters extracted*

Table II contains the relevant parameters determined as part of the PTC analysis; the key results include the excess noise factor of 1.268 for the iXON Life, lower than the maximum of $\sqrt{2}$. This is while the fixed-pattern noise was negligible, as is expected from the CCD architecture. It is interesting to compare these values to those similar ones reported in Table I—there is agreement on the order of magnitude, but not precisely.

**TABLE II.** Camera parameters derived from a PTC protocol for the cameras in 1 × 1 binning mode.

|  | Quest | Flash | iXON (300 gain) |
|---|---|---|---|
| $g$ | 9.35 | 2.17 | 142.9 |
| $F_n$ | 1 | 1 | 1.268 |
| $A_{FP}$ | 12.16 | 0.39 | 0 |
| $\sigma_{\text{read}}(e-)$ | 0.41 | 1.41 | 0.14 |

### *4. Noise maps*

The data collection procedure for the noise maps is similar to that for the PTC: a camera is placed in the uniform illumination set-up, where an ND filter is used to attenuate the light level to ∼10% of the camera's pixel well-depth.

The definition for the noise maps is relatively straightforward. Based on the captured stack of 100 frames imaged under identical conditions, the offset can be calculated with

$$\text{Off}_i = \text{Avg}_F\big[\text{DN}_{i,F}^0\big], \quad (A15)$$

while the dark signal non-uniformity has the following definition:

$$\text{DSNU} = \text{Std}_i\big[\text{Off}_i\big]. \quad (A16)$$

The read noise map has a similar definition,

$$\sigma_{\text{read},i} = \sqrt{\text{Var}_F\big[\text{DN}_{i,F}^0\big]}, \quad (A17)$$

while the read noise non-uniformity is given by

$$\text{RNNU} = \frac{\text{Std}_i\big[\sigma_{i,\text{read}}\big]}{\text{RMS}_{\text{read}}}, \quad (A18)$$

with RMS being the root-mean squared, which is computed as

$$\text{RMS}_{\text{read}} = \sqrt{\text{Avg}_{i,F}\big[\big(\text{DN}_{i,F}^0 - \text{Off}_i\big) - \sigma_{i,\text{read}}\big]^2}. \quad (A19)$$

Finally, the relative photoresponse can meanwhile be computed as

$$\text{rPR}_i = \frac{\text{Avg}_F\big[\text{DN}_{i,F}^S - \text{Off}_i\big]}{\text{Avg}_{i,F}\big[\text{DN}_{i,F}^S - \text{Off}_i\big]}, \quad (A20)$$

with the photoresponse non-uniformity (PRNU) being

$$\text{PRNU} = \frac{\sqrt{\text{Std}_i\big[\text{Avg}_F\big[\text{DN}_{i,F}^S\big]\big]^2 - \text{Std}_i\big[\text{Off}_i\big]^2}}{\text{Avg}_{i,F}\big[\text{DN}_{i,F}^S - \text{Off}_i\big]}. \quad (A21)$$

## APPENDIX B: BIOLOGICAL IMAGING

This appendix contains information relevant to the fluorescence microscopy system used and the preparation of biological samples for the purposes of imaging.

### 1. Light sheet microscope

The LSFM (see Fig. 9) in this study uses a femtosecond pulsed Ti:sapphire laser (Chameleon Vision S, Coherent), which has a tunable wavelength ranging from 690 to 1050 nm as an illumination source. The laser was tuned to 740 nm and the output passed through a telescope onto a scanning galvo mirror (Thorlabs); the reflection from which is focused onto the sample by the illumination objective (10× Nikon objective, 0.3 NA). In order to generate the sheet of light, the galvo mirror is scanned along the plane perpendicular to the detection objective; an arbitrary function generator (Keysight EDU33212A) is used to operate the galvo mirror with a 100 Hz ramp signal. The fluorescence excited by the light sheet is collected by a water dipping objective (40×, NA = 0.8, Nikon). Scattered excitation light is blocked with a shortpass filter while retaining the fluorescence signal. Finally, a tube lens forms an image of the emitted fluorescence ($f$ = 200 mm) at the plane of the camera. The detection arm of the microscope was shielded to ensure ambient light at the sample was minimized. Different cameras were switched into the camera port to capture images of the sample.

### 2. Sample preparation and imaging

To compare camera performance using autofluorescent samples, live mouse embryos at the blastocyst stage of development (96 h post-fertilization) were used. Embryos were mounted in culture media capable of maintaining a pH of 7.4 under atmospheric conditions.[12] During handling and imaging, embryos were maintained at 37 °C. The fluorophores FAD and NAD(P)H are the primary source of the fluorescent emission, both of which have high two-photon cross sections at the wavelength we used[44] and are known to be present within live mouse embryos.[4] Due to continued and dynamic development, imaging of embryos occurred within an 8 hour window of development to limit variability due to changes in cell proliferation/differentiation.

All the animal studies associated with this project were conducted in accordance with the Australian Code of Practice for the Care and Use of Animals for Scientific Purposes. The live mouse embryo samples were scavenged or obtained from studies conducted for other experimental purposes.

## APPENDIX C: POST-PROCESSING ALGORITHMS

In order to assess the effect of the algorithms using the methods outlined in Sec. III A, stacks of captured images were denoised, with each image within the stack being denoised independently.

For the ACsN algorithm, the MATLAB package supplied by the authors was used,[50] along with the camera calibration file for the generation of gain and offset maps; data collected as part of the PTC process were used for this purpose.

In contrast to ACsN, which utilized raw camera data, the other algorithms considered were entirely self-trained and, therefore, rely on sub-sampling to generate the training sets. Both algorithms were applied using Python.

The Neighbor2Neighbor algorithm employs neighbor sub-sampling, where the image, **x**, is first partitioned into $n \times n$ grids of size $k \times k$. Then, two neighboring pixels in each grid are randomly selected and sampled into two separate images, called $g_1$ and $g_2$. From here, $g_1$ is fed into the denoising network, a U-Net style architecture where we have set the depth to be 3, and is then trained for 100 epochs. The output of feeding $g_1$ into the network is then compared to $g_2$ to calculate a reconstruction loss value, $L_{rec}$. Following this, **x** is fed into the denoising network, where its output is sampled with neighbor sub-sampling to generate $h_1$ and $h_2$. A second loss is calculated with these image pairs, called the regularized loss, $L_{reg}$. A total loss that is the sum of $L_{rec}$ and $L_{reg}$ is calculated and used for back propagation to update the network parameters.

The Noise2Fast algorithm uses a sampling method known as checkerboard down-sampling, where the sampling mimics a checkerboard pattern. Given an image **x** of size $m \times n$, we can generate two sub-sampled images by assigning all even numbered pixels as one image and all odd numbered pixels as another, with each image being of size $m \times n$. These images are then either squeezed upward or to the left to generate $m \times \frac{1}{2}n$ and $\frac{1}{2}m \times n$ sized images, respectively. By doing this for both the odd-numbered and even-numbered pixel images, we obtain a total of four images. For training, a convolutional neural-network (CNN) is used, where left-squeezed images are considered as one set of training pairs, and up-squeezed images are considered as another. At the start of each epoch, a training pair is chosen at random and is fed into the denoising network. A binary cross-entropy loss is calculated and used to update the network parameters. The original unsampled image is then fed into the network, with its loss calculated and stored. The number of epochs is variable, as the network will continue to train until the loss of the original unsampled image shows no improvement after an additional 100 iterations.